# Efficient Metropolis-Hastings Proposal Mechanisms for Bayesian Regression Tree Models

M. T. Pratola*

October 23, 2013

**Abstract**

Bayesian regression trees are flexible non-parametric models that are well suited to many modern statistical regression problems. Many such tree models have been proposed, from the simple single-tree model to more complex tree ensembles. Their non-parametric formulation allows for effective and efficient modeling of datasets exhibiting complex non-linear relationships between the model predictors and observations. However, the mixing behavior of the Markov Chain Monte Carlo (MCMC) sampler is sometimes poor. This is because the proposals in the sampler are typically local alterations of the tree structure, such as the birth/death of leaf nodes, which does not allow for efficient traversal of the model space. This poor mixing can lead to inferential problems, such as under-representing uncertainty. In this paper, we develop novel proposal mechanisms for efficient sampling. The first is a rule perturbation proposal while the second we call tree rotation. The perturbation proposal can be seen as an efficient variation of the change proposal found in existing literature. The novel tree rotation proposal is simple to implement as it only requires local changes to the regression tree structure, yet it efficiently traverses disparate regions of the model space along contours of equal probability. When combined with the classical birth/death proposal, the resulting MCMC sampler exhibits good acceptance rates and properly represents model uncertainty in the posterior samples. We implement this sampling algorithm in the Bayesian Additive Regression Tree (BART) model and demonstrate its effectiveness on a prediction problem from computer experiments and a test function where structural tree variability is needed to fully explore the posterior.

*Department of Statistics, The Ohio State University, 1958 Neil Avenue, 404 Cockins Hall, Columbus, OH 43210-1247 (mpratola@stat.osu.edu). The author thanks Dr. Hugh Chipman for helpful feedback on an earlier draft of this manuscript. This work is dedicated to T. Zhang.

# 1 Introduction

Regression tree approaches to modeling complex nonlinear relationships have enjoyed increasing popularity in the statistical literature in recent years under the guise of Bayesian formulations (e.g. Chipman et al. (1998, 2002); Denison et al. (1998)). A variety of such Bayesian formulations have been developed, from single-tree models (Chipman et al., 1998, 2002; Denison et al., 1998), treed Gaussian Process models (Gramacy and Lee, 2008), sequential regression trees (Taddy et al., 2011) and Bayesian Additive Regression Trees (BART) (Chipman et al., 2010). Recent work also indicates that efficient and scalable parallel versions of regression tree models are possible (Pratola et al., 2013), which is timely given the explosion in the size and complexity of datasets today's applied statisticians are analyzing.

The benefits of regression tree models are well known; they allow for a flexible modeling approach that can handle a wide variety of nonlinear problems, have a simple and easy to understand structure, and offer fast predictive performance. At the same time, none of the drawbacks of conventional basis function approaches are present; for instance, no specification of a basis is required, and the computationally expensive matrix-algebra associated with basis function approaches is absent. The promise of the above Bayesian formulations of regression trees is to combine all of the benefits of regression tree models with all of the benefits of the Bayesian modeling paradigm, namely accounting for the various sources of uncertainty which are then propagated through to the posterior predictive distribution.

In practice, the aspirations of these various Bayesian regression tree methodologies are realized in many applications, but there are some problems that arise in certain cases. Primarily, it is well known that the Metropolis-Hastings (MH) proposals in the Markov Chain Monte Carlo (MCMC) sampler of these models can suffer from poor mixing (Wu et al., 2007), resulting in overfitting the data and under-representing model uncertainty. A few approaches that go some way towards mitigating this issue have appeared in the literature. The method of Taddy et al. (2011) approaches the problem by using a particle-based representation of the unknown posterior distribution. The BART model of Chipman et al. (2010) approaches the problem by forming a sum-of-trees representation of the data, where each tree is penalized to have a shallow depth. The idea in this approach is that with these shallow trees, the simple MH proposals that sometimes failed to explore the model space when used with deep trees will be adequate due to the vastly reduced search space. Another approach to improve mixing is the proposed "radical restructure" MH proposal developed in Wu et al. (2007). Their result suggested that mixing problems were completely eliminated when their restructure proposal was combined with the usual proposal mechanisms previously developed in the literature. However, their proposal is computationally expensive and does not scale well with high-dimensional problems (i.e. large number of covariates, $d$) due to the curse of dimensionality.

Our work was motivated by a situation where the BART model, even with the focus on shallow trees, exhibited poor mixing and severely underestimated posterior predictive uncertainty. The solutions we develop in this paper include a vastly more efficient version of the classical "change" proposal (Chipman et al., 1998; Denison et al., 1998) and a novel proposal mechanism that enables efficient searching of the tree space, thereby allowing the Markov chain to mix adequately leading to correct representation of model uncertainty. The proposed samplers are developed using the BART model and as such this paper will focus on that particular development and demonstrate the improvements by applying BART to simple examples to compare the performance of the old and new samplers. Nonetheless, the proposed samplers are applicable to Bayesian regression tree models in general.

In the next section, we motivate our development with a synthetic example from Wu et al. (2007) and a simple example from Computer Experiments (Sacks et al., 1989; Kennedy and O'Hagan, 2001; Oakley and O'Hagan, 2002; Higdon et al., 2008; Gramacy and Lee, 2008) where the existing BART sampler leads to inadequate mixing. In Section 3 we introduce our "perturb" move which is an enhanced version of the "change" proposal mechanism. In Section 4, we develop our novel tree rotation proposal mechanism. In Section 5, we apply the new BART samplers to the motivating problems to demonstrate the improvements realized. Finally, we conclude in Section 6.

## 2 Background And Motivating Examples

An acknowledged challenge of Bayesian regression tree models has been the difficulty to sometimes achieve proper mixing of the Markov chain. And, while this problem has been recognized since such models were established (Chipman et al., 1998; Denison et al., 1998), little progress has been made. Today, the majority of implementations continue to rely on the birth/death/change/swap proposals that were originally described. A notable exception to this is the work of Wu et al. (2007).

Regression trees model data using a stochastic binary tree representation that is made up of interior nodes, $T$, and a set of maps, $M$, associated with the terminal nodes. Since the tree is binary, any interior node, say node $\eta_i$, always has a left and right child, denoted as $l(\eta_i), r(\eta_i)$ respectively. Of course, all nodes except for the root node also have one parent node, $p(\eta_i)$. Frequently, it is common to refer to a node by its unique integer identifier $i$. For example, the root node $\eta_1$ is node 1. In this paper, we will sometimes also refer to a subtree starting at node $\eta_i$ simply as $T_i$. The example regression tree shown in Figure 1 summarizes our notation.

What does $T$ represent? Each internal node of a regression tree contains a split rule that depends on some covariate, and a split location, or "cutpoint". The representation $T$ is abstract, by which we mean that one might be referring to the tree $T$ or one might be referring to this modeling structure encoded in $T$. This modeling structure is simply the parameterization of the split rules at each node in the tree. Consider the $n \times d$ design matrix $X$ of covariates for our data. Each of the $d$ columns represents a covariate variable $v, v = 1, \ldots, d$ and each row $x$ corresponds to the observed settings of these covariates. Without loss of generality, assume that the covariates are scaled to the unit interval, so that $x_v \in [0, 1]$ and $x \in [0, 1]^d$.

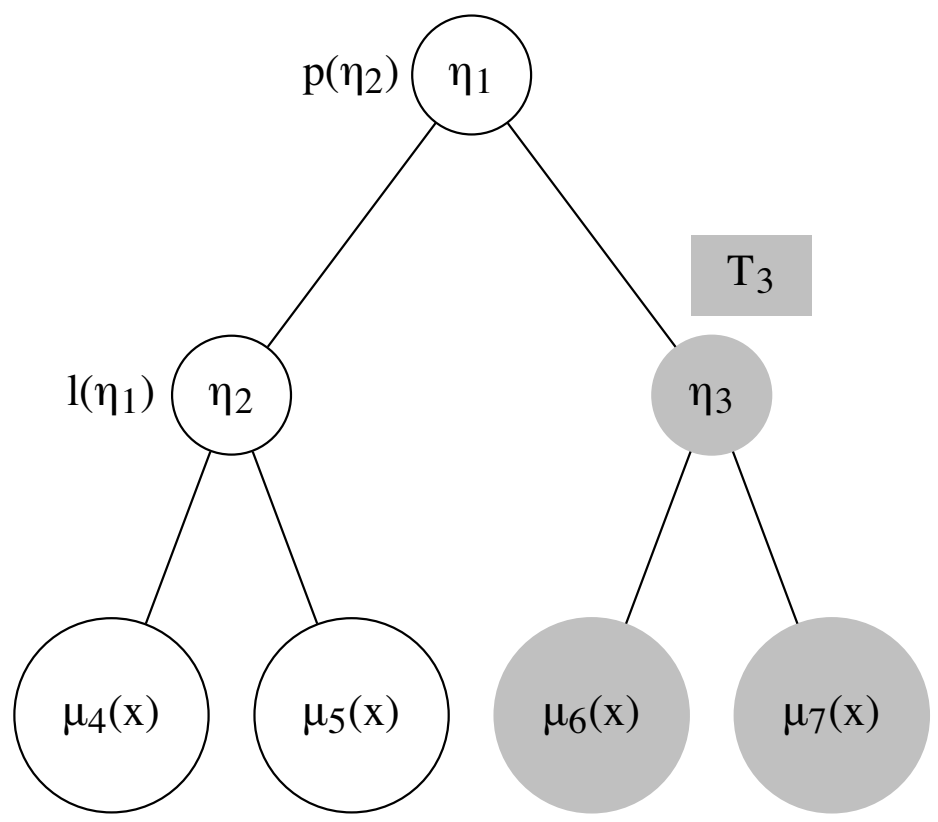

Figure 1: A sample binary regression tree. The tree is made up of internal nodes $T = \{\eta_1, \eta_2, \eta_3\}$ and has bottom-node maps $M = \{\mu_4(x), \mu_5(x), \mu_6(x), \mu_7(x)\}$. Each internal node has a rule of the form $v < c$ for some variable $v$ and some cutpoint $c$. When a rule is true for an arbitrary input $x$, we branch left, otherwise we branch right. Input $x$ maps to $\mu_5(x)$ if it branches left at $\eta_1$ and right at $\eta_2$. Note that node $\eta_2$ is also the left child of the root node, $l(\eta_1)$, while the root node $\eta_1$ is also the parent $p(\eta_2)$. The sub-tree $T_3$ denoted in grey consists of the node $\eta_3$ and all it's children.

Then the split rule at a given interior tree node is of the form $x_v < c$ which is parameterized by the chosen split variable $v$ and the cutpoint $c$.

The stochastic regression tree representation is enabled by treating the split variable $v$ as a discrete random quantity taking on values in $\{1, \ldots, d\}$ and treating the cutpoint $c$ as a discrete random quantity taking on values in $\{0, \frac{1}{n_v-1}, \ldots, \frac{n_v-2}{n_v-1}, 1\}$ where $n_v$ is the resolution of this discretization for variable $v$. For a continuous covariate, a choice of $n_v = 100$ is common, while for a discrete covariate this would be adjusted accordingly. In any case, the $n_v$'s are usually specified as fixed, known in the Bayesian formulation. Note that the modeling structure representation of $T$ could then be expressed as $T = \{(v_1, c_1), (v_2, c_2), \ldots\}$.

The regression tree is completed by specifying the maps at the terminal nodes. If there are $n_b = |M|$ terminal nodes, then we have the maps $M = \{\mu_1, \ldots, \mu_{n_b}\}$. These maps are simply functions that take as input the covariates $x$ that map to a given terminal node and produce a response $\mu_j(x)$. Common forms of the $\mu_j$'s are constants (i.e. $\mu_j(x) \equiv \mu_j$), a linear regression model, a Gaussian Process model (Gramacy and Lee, 2008), etc. Taken all together, the regression tree $T$ represents a partitioning of the covariate space $\chi$ and a mapping from an input covariate $x \in \chi$ to a response value encoded in $M$. To be more exact, the regression tree defines a function $g(x; T, M)$ which maps input $x$ to a particular response $\mu_j(x)$.

While there are many forms of regression tree models used in statistics, our work was motivated in particular by the Bayesian Additive Regression Tree (BART) model of Chipman et al. (2010). So, for the rest of this paper we use BART to implement and demonstrate our novel sampling strategies, although the methods developed are generally applicable.

## 2.1 BART

In the BART model, the idea is to represent the data $y$ as a sum of simple trees,

$$y(x) = \sum_{j=1}^{m} g(x; T_j, M_j) + \epsilon$$

where $\epsilon \sim N(0, \sigma^2)$. These simple trees are implemented by specifying constants for the terminal node maps and using a prior that penalizes the depth of each tree. The default number of trees used in this representation is $m = 200$, which seems to work well for a wide variety of problems. Conjugate normal priors on the terminal node $\mu$'s leads to a standard Gibbs sampler for the terminal node maps. Discrete uniform priors are placed on the split variables and split cutpoints and combined with a prior on the probability that a node is terminal leads to a Metropolis-Hastings algorithm for growing or pruning nodes in the tree. This growing and pruning are handled by aptly named birth and death proposals which either split a currently terminal node on some variable $v$ at some cutpoint $c$, or collapses two terminal nodes thereby removing a split.

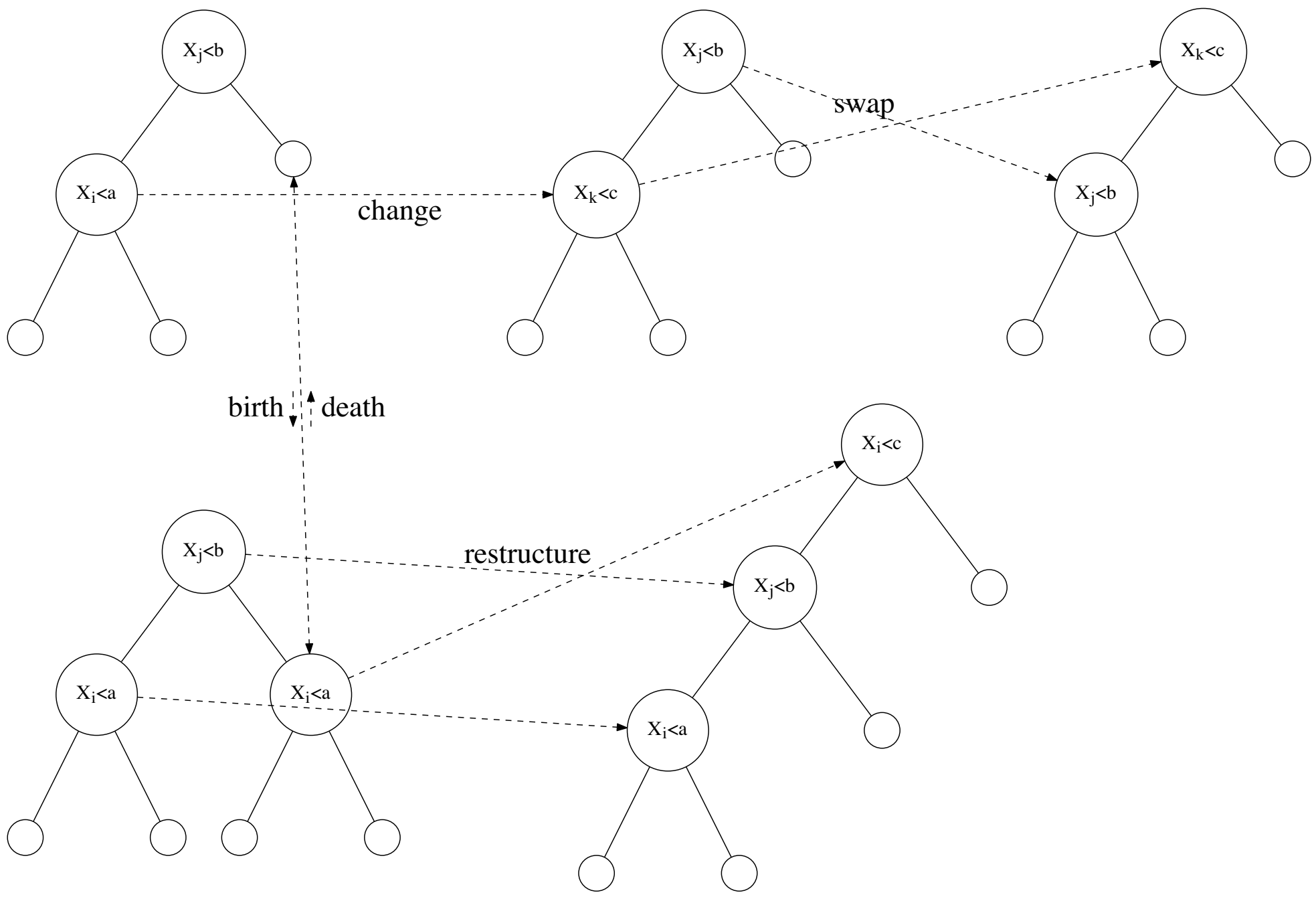

Figure 2: The birth, death, change and swap proposal mechanisms for fitting Bayesian regression trees via MCMC sampling. In addition, the restructure move introduced by Wu et al. (2007) is shown.

In addition to the birth/death MH steps, which allows the parameter space of the nodes at the bottom of the trees to be explored by the MCMC algorithm, there are additional change/swap proposals aimed at exploring the parameter space of nodes that are internal to the tree. These four MH proposal mechanisms were originally proposed in Chipman et al. (1998); Denison et al. (1998), and are summarized in Figure 2. For complete details of the MCMC algorithm for BART, the reader is referred to Chipman et al. (2010).

Using the four proposal mechanisms in single-tree regression models, it has frequently been found that the sampler initially mixes well for the first few iterations as the tree grows to fit the data, but then becomes trapped in a local mode being unable to accept death moves with any reasonable probability. At the same time, change/swap moves tend to have very low acceptance rates, further limiting the mixing that can occur. As a result, while the prediction of these fits can be quite good, there is a danger of over-fitting and the uncertainty bounds are often inaccurate.

One of the advantages of the BART model, with its additive representation of simple trees, was to facilitate easier acceptance of birth/death proposals because most of the trees would be shallow and only represent a small portion of the overall response signal. In addition, because of this shallow representation, removing the change/swap proposals was believed to be justified in Pratola et al. (2013). A similar simplification was also made in Taddy et al. (2011). This seems largely justifiable for many regression problems investigated, but we have since found that even the simple additive tree representation can suffer from poor mixing in certain problems. The applied problems where we have seen this occur come from computer experiments applications, where the measurement error variance $\sigma^2$ tends to be quite small and/or the dataset size is quite large. We show one such example next.

## 2.2 The Friedman Example

In computer experiments, a statistical emulator is used to model the outputs of simulators of complex physical processes (Sacks et al., 1989; Kennedy and O'Hagan, 2001; Oakley and O'Hagan, 2002; Higdon et al., 2008; Gramacy and Lee, 2008). To simulate such an example, we treated the deterministic Friedman function (Friedman, 2001) as if it were our simulator, sampled 5,000 observations from this simulator while adding i.i.d. normally distributed noise with variance $\sigma^2$ as in Pratola et al. (2013), and evaluated the use of BART as a flexible statistical emulator. When the measurement error was large, e.g. $\sigma^2 = 1$, the BART MCMC algorithm (with birth/death proposals only) was found to mix reasonably well, having an acceptance rate around 18% and the 90% credible interval having an empirical coverage of 81%. However, as the error variance was decreased, this behavior changed drastically.

The results of fitting BART when $\sigma^2 = 0.1$ are shown in Figure 3. For this example, there were again $n = 5,000$ observations generated from the Friedman function at random settings of the covariates, with i.i.d. Normal noise added to form the observation. The BART model was fit with $m = 200$ trees and allowed to burn-in for 5,000 iterations and a further 5,000 iterations were drawn as samples from the posterior.

The plot of the change in log integrated likelihood from the birth/death proposals is shown for all $10,000$ iterations of the MCMC and all $m = 200$ trees. In this case, one can see that the changes in log integrated likelihood from birth/death proposals are all large negative values, save for the first few proposals at the beginning of the run. This is reflected in the very low acceptance rate that is just around 4%. In effect, the tree structure became stuck in a local mode with, for all practical considerations, zero chance of moving to a different area of tree-space that could give an equally good fit to the data.

The empirical coverage of the 90% credible interval shown in the figure is also very low at around 53%. Since the tree structure of the model is not being explored by the sampler in this example, it suggests that the uncertainty coming from the terminal node $\mu$'s is only accounting for roughly half of the true uncertainty that should be explored by the MCMC sampler. This missing uncertainty is attributed to variability in partition rules of the interior tree nodes and other structural variability of the regression trees in this example.

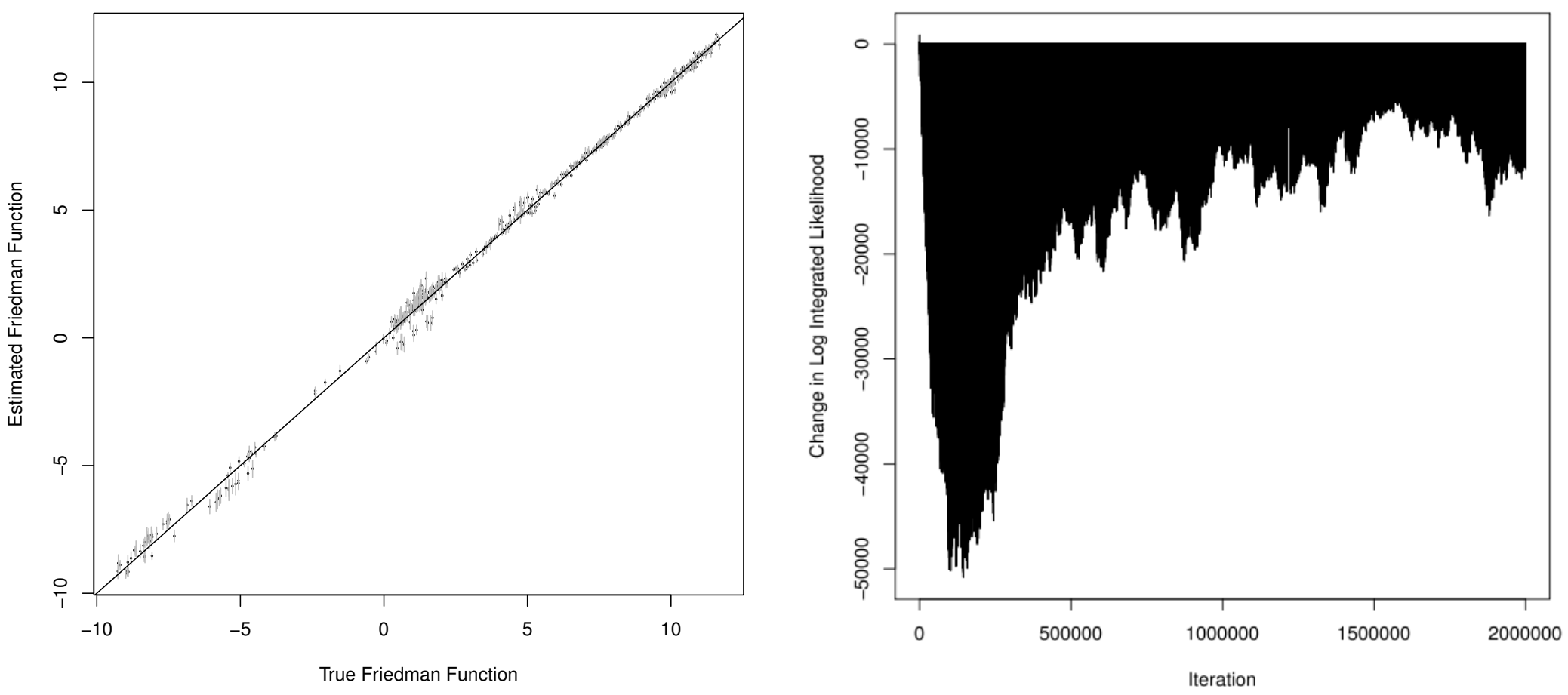

Figure 3: Credible intervals for posterior predictions of the Friedman function with $\sigma^2 = 0.1$ (left) and the change in log integrated likelihood for the birth/death MH proposals (right). The credible intervals under-represent the uncertainty and the very low acceptance rate of birth/death proposals of 4% indicate poor mixing of the MCMC sampler.

## 2.3 A Synthetic Example

This example is taken from Wu et al. (2007), as it serves as a simple demonstration where proper mixing of the regression tree structure is important. The synthetic dataset has $p = 3$ covariates and the response $y$ was calculated at $n = 300$ settings of these covariates as:

$$y(x) = \begin{cases} 1 + N(0, 0.25) & \text{if } x_1 \le 0.5 \text{ and } x_2 \le 0.5 \\ 3 + N(0, 0.25) & \text{if } x_1 \le 0.5 \text{ and } x_2 > 0.5 \\ 5 + N(0, 0.25) & \text{if } x_1 > 0.5 \end{cases} \tag{1}$$

The covariates were generated as $x_{1i} \sim \text{unif}(0.1, 0.4)$ for $i = 1, \ldots, 200$, $x_{1i} \sim \text{unif}(0.6, 0.9)$ for $i = 201, \ldots, 300$, $x_{2i} \sim \text{unif}(0.1, 0.4)$ for $i = 1, \ldots, 100$, $x_{2i} \sim \text{unif}(0.6, 0.9)$ for $i = 101, \ldots, 200$, $x_{2i} \sim \text{unif}(0.1, 0.9)$ for $i = 201, \ldots, 300$, $x_{3i} \sim \text{unif}(0.6, 0.9)$ for $i = 1, \ldots, 200$ and $x_{3i} \sim \text{unif}(0.1, 0.4)$ for $i = 201, \ldots, 300$. Note that with this function and the covariates generated as described, the effects of $x_1$ and $x_3$ are confounded.

We fit BART to this dataset using only $m = 1$ trees and found that the acceptance rate of tree moves (after the initial few steps of the sampler) was 0. In effect, the sampler collapsed on a single tree representation of the data and would not accept any birth/death proposal that might lead to an alternative representation of the data. The tree that was found is the 4-node representation in the top-right of Figure 10. If one were to blindly believe this fit to the data, it would appear that $x_3$ has no effect on the response, whereas in fact we should conclude that either $x_1$ or $x_3$ (or both) may affect the response.

# 3 Perturbation Proposal

Besides the birth/death proposals, the change proposal has been a popular MH move for exploring the posterior of Bayesian regression tree models. This move can be thought of as changing a cutpoint, a variable or both simultaneously. We prefer to take a simple one-at-a-time approach to our sampling algorithm, and so will first focus on proposing a new cutpoint at internal node $i$, denoted $c_i$, given the cutting variable $v_i$ and the rest of the tree structure $T$.

Interestingly, while the "change" proposal has been around for some time, there are some varied explanations of its implementation in the literature. Denison et al. (1998) draw from a uniform proposal distribution on the range of values $v_i$ takes. This amounts to an independence sampler when the prior is uniform with endpoints taken as the min and max observed covariate values in the dataset. Chipman et al. (2002) take a similar approach, drawing proposals from the prior distribution. In Chipman et al. (1998), the proposal distribution is restricted to functions that depend only on the part of $T$ ancestral to node $i$. Their default is again to take the proposal to be uniform, but this time restricted by the requirement that values drawn from this proposal cannot lead to empty terminal nodes. This requires propagating the data through the proposed tree to ensure this condition is met, which is an expensive operation. Wu et al. (2007) and Chipman et al. (2010) use the above approach in their change proposal while Gramacy and Lee (2008) use a random walk approach, essentially incrementing or decrementing the cutpoint by the single smallest increment available from the covariate matrix $X$.

It is useful to think about the meaning of the cutpoint and split variable components of regression tree models. Namely, the cutpoints of a regression tree relate to the "wiggliness" of the response being modeled while the split variables are indicative of the importance of that variable in modeling, or explaining the variability, of the response.

For a given collection of split variables and cutpoints $\{v_i, c_i\}$ that form a fitted regression tree model, this collection represents a discrete encoding of the model. Changing just a single $v_i$ or a single $c_i$ represents an entirely different model. However, intuitively, changes involving $c_i$ are in some sense more local than changes involving $v_i$.

A reasonable proposal for a particular $c_i$ conditional on everything else is an interval $(a_i^{v_i}, b_i^{v_i})$ that should take into account the full tree structure $T \setminus c_i$. For node $i$, let $C_{p(i)}^{v_i}$ be the collection of cutpoints for all nodes ancestral of node $i$ that split on variable $v_i$, and let $C_{l(i)}^{v_i}$ (similarly $C_{r(i)}^{v_i}$) be the collection of cutpoints for all nodes in the left subtree of node $i$ (similarly right subtree) that split on variable $v_i$.

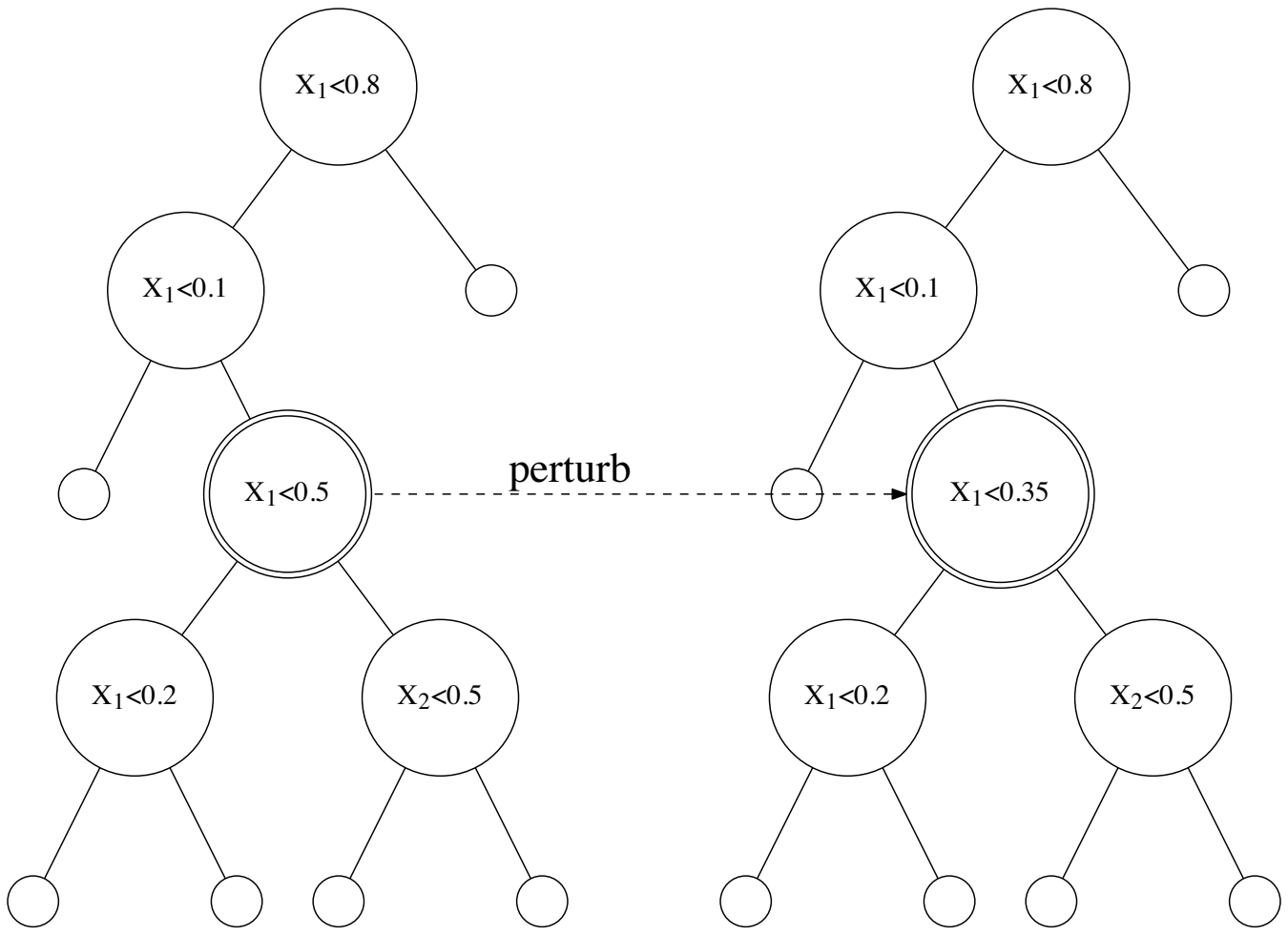

Figure 4: An example of a perturb proposal at node 5. In this case, we have $C^{v_1}_{p(5)} = \{0.1, 0.8\}, C^{v_1}_{l(5)} = \{0.2\}$ and $C^{v_1}_{r(5)} = \{\}$, giving the open interval $(a^{v_1}_5, b^{v_1}_5) = (0.2, 0.8)$ from which to draw proposals for a new cutpoint value.

A uniform proposal that is consistent with the full tree structure draws a cutpoint uniformly from the interval

$$(a^{v_i}_i, b^{v_i}_i) = \left( max\left(0, min(C^{v_i}_{p(i)}), max(C^{v_i}_{l(i)})\right), min\left(1, max(C^{v_i}_{p(i)}), min(C^{v_i}_{r(i)})\right)\right). \quad (2)$$

Since the dimensionality of the model does not change with such proposals, the MH ratio involves likelihoods rather than integrated likelihoods. Note that the proposal does not directly depend on the data and so is easily computed. Also, making use of the same restriction on a minimum number of observations in terminal nodes as used in birth/death proposals, the move can be rejected if the minimum requirement is not met during calculation of the likelihood.

A simple example of the perturb proposal is shown in Figure 4, where we try to perturb the cutpoint at node 5 which has initial rule "$x_1 < 0.5$". If we generate proposals from the prior on the interval $(0, 1)$, immediately 40% of these proposals are outside the valid range $(a^{v_1}_5, b^{v_1}_5) = (0.2, 0.8)$ and will be rejected. These rejections occur solely because some terminal nodes will be unreachable and will therefore contain no data, which automatically leads to rejection by the MH sampler. For example, proposing the cutpoint to be $c_5 = 0.05$ would be rejected because no data would be mapped to the right sub-tree of the node

"$x_1 < 0.2$" due to the constraint "$x_1 < 0.1$" at node 2. Or, proposing the cutpoint to be $c_5 = 0.9$ would be rejected because no data would map to the right sub-tree of node 5 due to the constraint "$x_1 < 0.8$" at node 1. Similarly, generating proposals from the prior conditioned on the structure of the tree above node 5 (i.e. from $(0.1, 0.8)$) leads to about 17% of such proposals being outside the valid range and therefore guaranteed to be rejected. For instance, proposing $c_5 = 0.15$ would be rejected because no data would map to the right sub-tree of the node "$x_1 < 0.2$" (this would require both $x_1 < 0.15$ and $x_1 > 0.2$, which is clearly not possible). In contrast, the perturb proposal generates cutpoints from the valid range $(a_5^{v_1}, b_5^{v_1}) = (0.2, 0.8)$, which accounts for constraints from the ancestral (i.e. $C_{p(5)}^{v_1}$) and descendant (i.e. $C_{l(5)}^{v_1}, C_{r(5)}^{v_1}$) parts of the tree about node 5 to avoid any such spurious rejections.

The proposal can be made more flexible by introducing a multiplicative constant $\alpha \in (0, 1]$ (i.e. independent of node), or $\alpha_v \in (0, 1])$ (i.e. a unique scaling for each of the covariate variables), that scales the proposal to be more local, and the parameter $\alpha$ can be updated in an adaptive MCMC framework to target higher acceptance rates if needed. This approach chooses cutpoints that lie in the scaled interval,

$$\left( max\left( c_i - \alpha \left( \frac{b_i^{v_i} - a_i^{v_i}}{2} \right), a_i^{v_i} \right), min\left( c_i + \alpha \left( \frac{b_i^{v_i} - a_i^{v_i}}{2} \right), b_i^{v_i} \right) \right),$$

where the $a_i^{v_i}, b_i^{v_i}$ are computed as in (2), thereby making the proposals in a more local neighborhood about the current cutpoint $c_i$ within the valid interval $(a_i^{v_i}, b_i^{v_i})$. If we instead fix $\alpha$ to be very small, forcing minute local moves in cutpoint values, then the proposals behavior would be similar to that of Gramacy and Lee (2008). In practice, we have fixed $\alpha$ to cover 85% of the range defined in the interval (2), which has worked well in the problems we have tried. In general, the method does not seem very sensitive to reasonable settings of $\alpha$ in our experience.

## 3.1 Perturb Within Change-of-Variable

While the perturbation proposal has the clear interpretation of exploring the "wiggliness", or spatial variability of the response being modeled, it is less intuitive what variability the usual change-of-variable proposal explores. Changing a variable at some internal node within a regression tree implies that for all observations mapping to that internal node, the variability of the response is well represented by the tree structure below this internal node no matter if the node splits on the current variable or the variable we propose to change to. This could be reasonable if

(i) the partition of observations when using the new variable is unique from the partition of observations using the current variable, but nonetheless yields two partitions that are adequately explained by the tree structure below the current node , or,

(ii) the partition of observations is nearly the same using either variable, which occurs when the two covariates under consideration are dependent on one another, such as the highly correlated covariates found in the synthetic example of Section 2.3.

In our opinion, the latter case seems more likely and so we develop a change-of-variable proposal that explores the variability in the posterior that results from such correlated covariates. The approach taken is essentially to precondition the change-of-variable proposal by taking into account the empirical correlation structure of the covariate matrix.

Using the pairwise correlation between current variable $v_k$ and other variables $v_{\backslash k}$, a change of variable at node $i$ is proposed in the following way. First, the probability of changing $v_k$ to $v_j, j \neq k$ should be relatively higher if the current variable is highly correlated with another (or many other) variables in the dataset. On the other hand, this probability should be very small if the current variable is relatively independent (uncorrelated) with all other variables. This is accomplished by proposing a transition from $v_k$ to $v_j$ with probability proportional to

$$\frac{Cor(X_k, X_j) \times \mathcal{I}_{\left(a_i^{v_j}, b_i^{v_j}\right) \neq \{\}}}{\sum_l Cor(X_k, X_l) \times \mathcal{I}_{\left(a_i^{v_l}, b_i^{v_l}\right) \neq \{\}}} \tag{3}$$

where $Cor(\cdot, \cdot)$ represents the sample correlation estimated from the full dataset in the usual way. Here the indicator functions ensure that we only give positive probability to transition to variables which have cutpoints available at the node in question. That is,

$$\mathcal{I}_{\left(a_i^{v_l}, b_i^{v_l}\right) \neq \{\}} = \begin{cases} 1 & \text{if there are cutpoints available at node } i \text{ for variable } v_l \\ 0 & \text{otherwise} \end{cases}$$

Note that if $v_k$ is independent of all other variables, then with high probability this formula will propose staying at variable $v_k$. If $v_k$ is highly correlated with a single other variable $v_j$, then this formula will lead to proposals that stay at $v_k$ about 50% of the time and propose transitions to $v_j$ about 50% of the time, and so on.

This procedure could be made more flexible by only calculating the correlations using the $x$'s that map to node $i$. This would then be taking into account more local information which may be useful when the relationship between covariates is more complex, such as when modeling waterways or other geographically constrained responses (e.g. Rathbun (1998); Løland and Høst (2003); Pratola (2006)). In such cases, using only the $x$'s mapping to node $i$ would amount to a locally linear approximation of the relationship between covariates.

On the other hand, for nodes near the bottom of the tree, using only $x$'s mapping to a near-bottom node $i$ may lead to having a very small number of $x$'s mapping to node $i$ and if the covariate space is relatively large then the procedure is in some sense underdetermined and sensitive to the small number of observations mapping to $i$. Hence, in practice we have used the full sample correlations in forming our change-of-variable proposal procedure.

A final matter to note is the implementation of this preconditioner when the covariate dimension is large. In such instances, many spurious small correlations between variables may appear. We suggest treating such situations by using an empirical cutoff (e.g. all sample correlations $\leq 0.30$ are replaced with 0), although other approaches are also feasible.

# 4 Tree Rotation Proposal

A limitation with perturb and change-of-variable proposals is that they do not directly explore radically different tree arrangements nor do they change the dimension of the tree itself. In fact, of all the proposal mechanisms that have been developed in the literature, only the birth/death move changes dimensionality of the model. Because these moves can only alter the very bottom of the tree, it is very unlikely in a practical amount of time for a regression tree MCMC algorithm to fully explore the space of nearly equivalent trees that have high posterior probability.

Here we will develop a more radical proposal, which can be thought of as a multivariate generalization of the simple univariate rotation mechanism found in the binary search tree literature. Our tree rotation proposal allows dimension-changing proposals to occur at any interior node of a tree, and directly moves between modes of high posterior probability through such moves. The basic idea of our rotate algorithm is demonstrated in Figure 5, which shows one possible trajectory of tree arrangements that can be constructed through two rotation moves applied to the same node.

In the example of Figure 5, the rotation algorithm can be explained in a series of sequential steps. For instance, to perform the right-rotation at the node "$X_i < a$" given the current tree structure found in the top-left of Figure 5, the steps are:

1. Swap the rule of the node "$X_i < a$" with the rule of its parent "$X_j < b$". We now have the rule $X_j < b$ at the rotation node and rule $X_i < a$ at the parent node.

2. Create a new right-child node for the parent node and also assign it the rule "$X_j < b$". Attach sub-tree $T_s$ to be the right-child of this new node, and its left child is now $T_r$. The right child of the rotation node is changed from $T_r$ to a new copy of $T_s$.

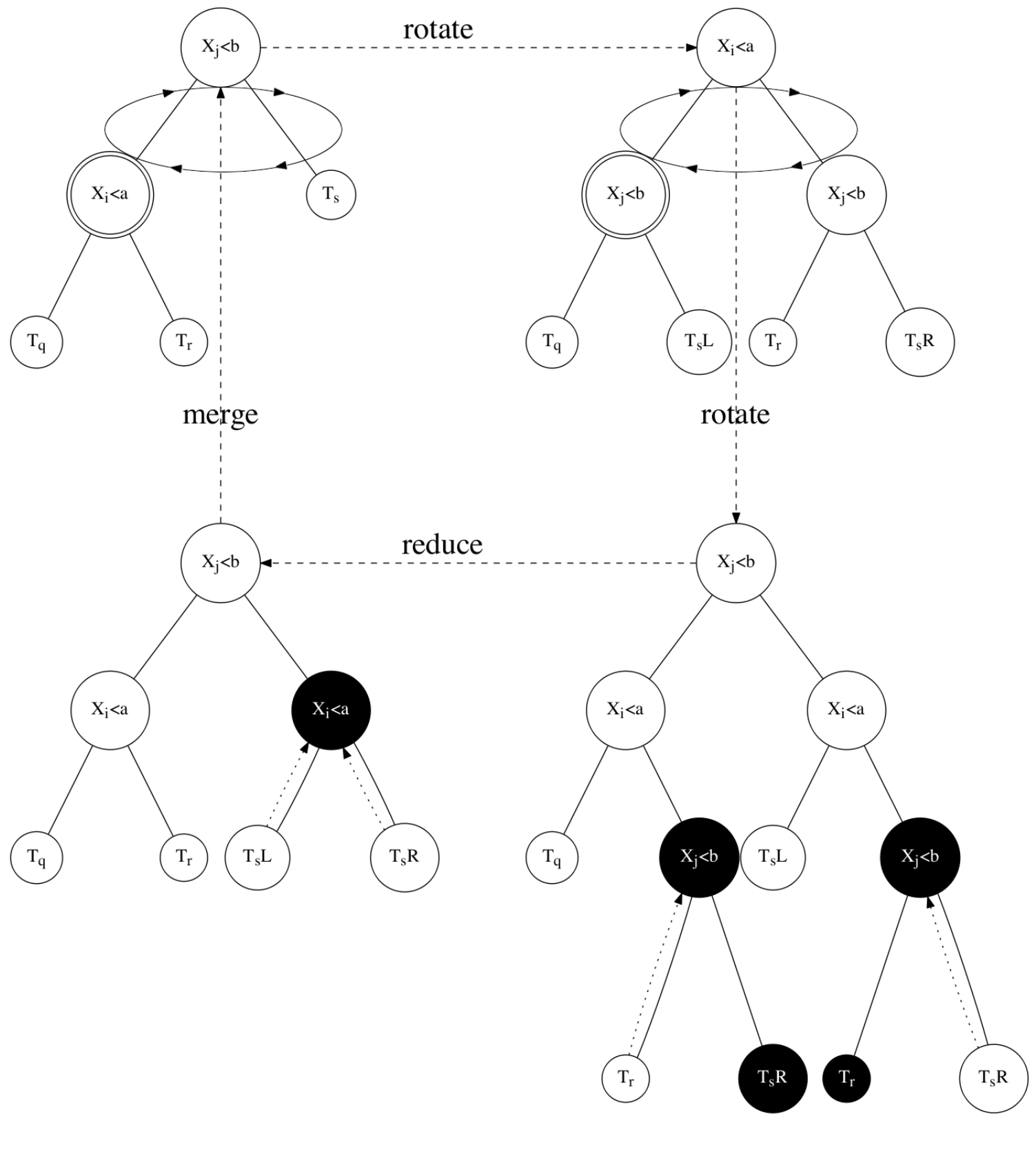

Figure 5: Two rotate moves applied sequentially at the same node (shown as double circled). For simplicity, the sub-trees $T_q, T_r$ are assumed not to split on $X_i, X_j$ and/or could simply be bottom-nodes, while $T_s$ is any arbitrary sub-tree. Note that the rotate operation is undone by a subsequent rotate when the model state has remained unchanged since the initial rotate. In the second rotate move, inadmissible sub-branches are cut and a merge operation is performed in order for the proposed tree to lie in the set of all possible tree models that can be reached by the basic birth/death process.

3. Remove duplicate "$X_i < a$" rules that may occur at the first node of both copies of tree $T_s$ (there are none in the first rotation, but this updating occurs in the subsequent rotation, see the bottom-right in Figure 5).

4. Divide (or "cut") both copies of sub-trees $T_s$ along the $X_i < a$ rule. This creates the distinct modifications $T_s^L$ and $T_s^R$ that appear in the top-right of Figure 5.

5. Merge the left and right subtrees of the new right-child node. In the first rotation, no merging is possible, but in the subsequent rotation, merging reconstructs $T_s$ from $T_s^L$ and $T_s^R$ as shown in the bottom-left of Figure 5. When this merging is possible, the new right node is deleted and the parent's right child again becomes $T_s$ as recovered by the merging operation.

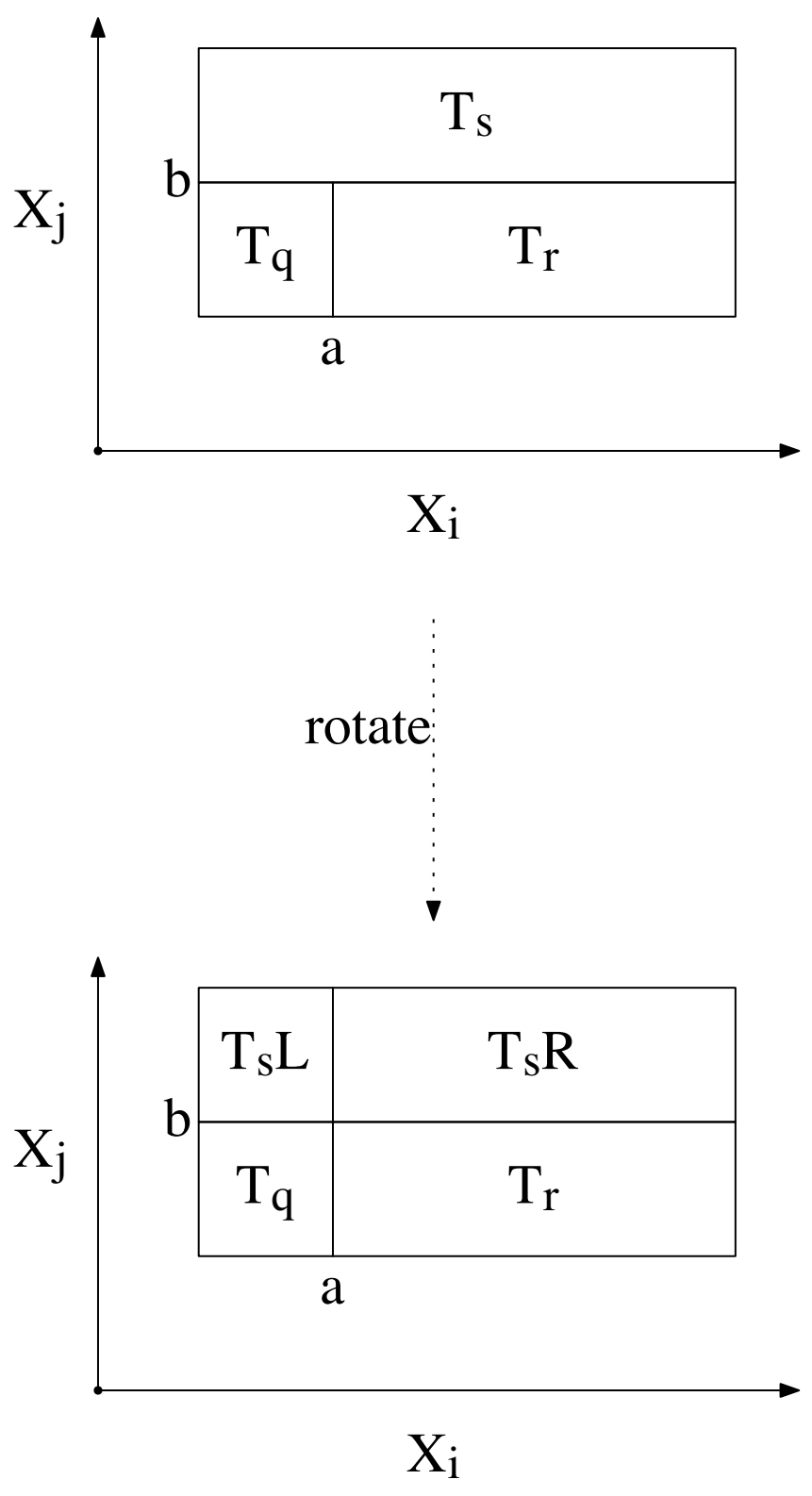

Figure 6: Effect of the first rotation in Figure 5 when viewed in the $X$-plane. Note that the rotation effectively extends the $X_i < a$ rule across the $X_j < b$ boundary. This causes the original tree $T_s$ to be split into $T_s^L$ and $T_s^R$. However, the actual decomposition of the $X - space$ has not otherwise changed between the original and rotated trees. It is only upon drawing new bottom-node $\mu$'s and/or subsequent birth/death or perturb proposals that change the prediction of the tree. This motivates how the rotate proposal moves between modes of high posterior probability.

The notion that a rotate traverses between modes of high posterior probability is motivated by viewing the effect of the first rotation performed in the example of Figure 5 in the plane of the $X$-space that is being affected, as shown in Figure 6. In this figure, we see that the rotation has extended a rule further through the covariate space at this level of the tree. Note that the rotation allows for the possibility of the same mapping of $x's$ to bottom-node $\mu$'s after the rotate. For example, if $T_s$ does not split on $X_i$ or $X_j$ then we will have $T_s^L = T_s^R = T_s$. Indeed, if the $\mu$'s were fixed, the rotated tree would return the exact same prediction as the original tree whether $T_s$ splits on $X_i, X_j$ or not.

Another way to view the behavior of a rotation is how $x$'s can map to the sub-trees $T_q, T_r$ and $T_s$. Before the initial rotation, the set of $x$'s that map to sub-tree $T_q$ must satisfy $\{X_j < b\} \cap \{X_i < a\}$ and those that map to sub-tree $T_r$ satisfy $\{X_j < b\} \cap \{X_i > a\}$. After the first rotation, it is easily verified by inspecting Figures 5 and 6 that these mappings remain unchanged. Whereas, before rotation the $x's$ mapping to $T_s$ must satisfy $\{X_j > b\}$ (*irregardless* the value of $X_i$) while after the mapping we have that the $x$'s that map to $T_s^L$ satisfy $\{X_j > b\} \cap \{X_i < a\}$ and those mapping to $T_s^R$ satisfy $\{X_j > b\} \cap \{X_i > a\}$. Note, however, that $\{\{X_j > b\} \cap \{X_i < a\}\} \cup \{\{X_j > b\} \cap \{X_i > a\}\} = \{X_j > b\}$, the original mapping for $T_s$.

These described features of the rotation operation allow us to satisfy the seemingly contradictory needs of a structural tree proposal that (i.) is a local and computationally feasible operation, (ii.) allows the sampler to move between very different tree arrangements of differing dimensionality, and yet (iii.) moves directly between tree arrangements that have high posterior probability.

## 4.1 The Rotation Operator, $\mathcal{R}$

More formally, for a rotatable node $\eta_i$ (i.e. an interior node) who is the left child of $p(\eta_i)$, a right-rotation proposal $T' = \mathcal{R}T$ is constructed according to the pseudo-code shown in Listing 1 of Appendix A (a similar algorithm for a left-rotation applies if $\eta_i$ is the right child of $p(\eta_i)$). Note that a rotatable node is simply an internal node of the tree except for the root node, since left/right rotation at the root node is equivalent to rotation at its left/right children.

While the details of the rotation can be found by studying the pseudo-code in Appendix A, we can more succinctly summarize these details by viewing the rotation operator as the composition of simpler operations, that is,

$$\mathcal{R}T = \mathcal{R}^L_{merge}\mathcal{R}^R_{merge}\mathcal{R}^L_{cut}\mathcal{R}^R_{cut}\mathcal{R}^R_{rot}T$$

where $\mathcal{R}_{rot}$ sets up the initial re-arrangement of the tree structure for the rotation, $\mathcal{R}_{cut}$ performs the cut operations on $T_s$ and $\mathcal{R}_{merge}$ performs the merge operations. We describe each of these in further detail next.

### 4.1.1 $\mathcal{R}^R_{rot} \equiv \mathcal{R}^R_{rot}(\eta_i; T)$

First, the operation $\mathcal{R}^R_{rot}(\eta_i; T)$ does the initial setup of the right-rotation at node $\eta_i$ of tree $T$ (a left rotation, $\mathcal{R}^L_{rot}$ would be similar). For instance, starting from the top-left arrangement in Figure 5, this operation swaps the rules of $\eta_i$ and p($\eta_i$), and also introduces a new node for r(p($\eta_i$)) also using the rule that was in $p(\eta_i)$. At the same time, the sub-tree $r(\eta_i)$ is moved to become the subtree $l(r(p(\eta_i)))$.

Subsequently, the sub-trees $r(\eta_i)$ and r(r(p($\eta_i$))) are initialized to duplicates of $T_s$, where $T_s$ is the sub-tree $r(p(\eta_i))$ in the original tree $T$.

#### 4.1.2 $\mathcal{R}^L_{cut} \equiv \mathcal{R}^L_{cut}(r(\eta_i), v_{p(\eta_i)}, c_{p(\eta_i)}; T)$, $\mathcal{R}^R_{cut} \equiv \mathcal{R}^R_{cut}(r(r(p(\eta_i))), v_{p(\eta_i)}, c_{p(\eta_i)}; T)$

Since both of the right subtrees $r(\eta_i)$ and $r(r(p(\eta_i)))$ of the tree are now under the constraint of the new rule appearing in $p(\eta_i)$, they need to be made consistent with this rule. This is performed by the cutting operations which remove inadmissible sub-subtrees splitting on variable $v_{p(\eta_i)}$, leading to the modified $T^L_s$ and $T^R_s$ as shown in the top-right arrangement of Figure 5. Pseudo-code describing the left-wise cut operation is shown as Listing 2 in Appendix B (a similar procedure performs right-wise cutting). In essence, the rotation operation leads to "dividing" the tree $T_s$ along level $c_{p(\eta_i)}$ of variable $v_{p(\eta_i)}$. Accordingly, sub-subtree $T^L_s$ is arrived at by removing all nodes from $T_s$ that do not satisfy the rule $v_{p(\eta_i)} < c_{p(\eta_i)}$ while sub-subtree $T^R_s$ is arrived at by removing all nodes from $T_s$ that do not satisfy the rule $v_{p(\eta_i)} > c_{p(\eta_i)}$.

#### 4.1.3 $\mathcal{R}^L_{merge} \equiv \mathcal{R}^L_{merge}(l(\eta_i), r(\eta_i), v_{\eta_i}, c_{\eta_i}; T)$, $\mathcal{R}^R_{merge} \equiv \mathcal{R}^R_{merge}(l(r(p(\eta_i))), r(r(p(\eta_i))), v_{p(r(\eta_i))}, c_{p(r(\eta_i))}; T)$

In a subsequent rotation, the "dividing" of tree $T_s$ can be undone so that $Prob(\mathcal{RR}T = T) > 0$, as required for a valid MCMC algorithm. This is handled by the merge operations, which have the effect of going from the arrangement in the top-right of Figure 5 back to the original arrangement with some probability. Note that you can't be both cutting and merging the sub-tree $T_s$ on a rotate proposal. If you are merging, then the transition probability $P(T \to T')$ needs to take into account that the probability of moving from $T$ to $T'$ is affected by the number of ways that two subtrees can be merged into the new tree. For example, suppose we are attempting to merge $T^L_s$ and $T^R_s$, and say we are at node 2 in both trees which we will call $\eta^L_2, \eta^R_2$ respectively. If both nodes split on $v_i$ (i.e. $v^L_2 = v^R_2 = v_i$) then clearly it must be the case that $c^L_2 < c_i$ and $c^R_2 > c_i$ and that in the original tree these two nodes were a parent-child pair. But, the original tree could have had $\eta^L_2$ as the parent with $\eta^R_2$ as the right-child or $\eta^R_2$ as the parent with $\eta^L_2$ as the left-child. In performing the merge operation, one of these reconstructions need be chosen randomly, and the total number of such random decisions is recorded so that the accept/reject calculation can be correctly calculated. In Listing 1, these counts are stored in the $n^1_m, n^2_m$ variables. These two count variables reflect the fact that the merging operation need be applied to $l(\eta_i) + r(\eta_i)$ merged along $v_{\eta_i}, c_{\eta_i}$ and to $l(r(p(\eta_i))) + r(r(p(\eta_i)))$ merged along $v_{p(r(\eta_i))}, c_{p(r(\eta_i))}$ where '+' represents merging.

On the other hand, if we are not merging then we (may) be cutting. If we are cutting then we need to adjust the return transition probability $P(T' \to T)$ to take into account that the probability of returning to the current configuration from the rotated configuration $T'$ is again affected by the number of ways that two sub-trees in the rotated tree can be merged in the inverse step to return to the current tree. In Listing 1, these counts are stored in the $n^1_s, n^2_s$ variables.

In calculating the number of merges possible in both the forward proposal and in calculating the probability of inverting the rotation, one need recognize that a non-trivial merge of both $l(\eta_i)+r(\eta_i)$ and $l(r(p(\eta_i)))+r(r(p(\eta_i)))$ leads to an inadmissible state. A non-trivial merge is one that does not retain the original variable and cutpoint node (e.g. $v_{\eta_i}, c_{\eta_i}$ or $v_{p(r(\eta_i))}, c_{p(r(\eta_i))}$). If both merges are non-trivial, then the (variable, cutpoint) pair required for inverting the rotation has been removed from the tree which would not allow inversion to take place. This can be seen in the visual summary of trajectories created by the rotate, cut and merge process for an initial right-rotation shown in Appendix D.

## 4.2 Calculating the M-H Ratio

Note that in calculating the log-integrated likelihoods for the accept/reject step, one need only consider terminal nodes that are part of the sub-tree of $p(\eta_i)$ since the rest of the tree $T$ remains unchanged by the rotation proposal. Hence, the computational cost of a rotation step, while greater than a simple birth/death proposal, is much reduced compared to a more drastic restructure move such as the proposal of Wu et al. (2007).

While the operation of cutting unreachable sub-branches is entirely deterministic, there are usually many possible merging arrangements and the stochasticity of the rotation proposal comes from selecting one of these arrangements at random. If one considers all permutations of variable and cutpoint rules, the list of merges is quite lengthy but fortunately most are inadmissible. The actual number of possible merges at any given level of the left and right trees are shown in Figure 7. Furthermore, many of these arrangements lead to identical merges, yielding only 7 unique merge types at a given level of the left and right trees.

The actual number of merge reconstructions can, of course, be higher due to the recursive definition of the merging operation. For example, the merge type itemized as #7 in the second column of Figure 7 is shown in Figure 8. This example demonstrates the recursive nature of the merging operation, and this recursion is followed down the levels of the right and left trees when counting the number of possible merges in the proposal distribution. We explicitly recursively calculate these counts when constructing a rotation proposal. The remaining six unique merge types are shown in Appendix C.

A final item to note in calculating the acceptance ratio is determining the probability of rotating at node $\eta_i$ in both the forward proposal and the inverse step. For the forward proposal, this is just the reciprocal of the number of internal nodes (less the root node) in the tree. However, for the inverse step it may be possible to rotate back to the current configuration from both $\eta_i$ or the node "opposite" of $\eta_i$ (which is $r(p(\eta_i))$ in a right-rotation). In this case, the probability is 2 over the number of internal nodes in the rotated configuration.

Taking all these considerations into account, the acceptance ratio can be calculated as

$$\alpha = min\left(1, \frac{\pi(T')p_r(T')p_s^1 p_s^2 L(T')}{\pi(T)p_r(T)p_m^1 p_m^2 L(T)}\right) \tag{4}$$

| Arrangement | Merge Type | l->v=r->v | l->c=r->c | l->v=vi | r->v=vi | l is leaf | r is leaf |
|---|---|---|---|---|---|---|---|
| 1 | 1 | x | x | x | x | | x |
| 2 | 2 | x | x | x | x | x | |
| 3 | 3 | | | | | x | x |
| 4 | 4 | x | x | | | | |
| 5 | 5 | x | x | | | | x |
| 6 | 5 | x | x | | | x | |
| 7a | 5 | | | | | | |
| 7b | 5 | | | | | | x |
| 7c | 5 | | | | | x | |
| 8a | 5 | | | x | | | |
| 8b | 5 | | | x | | | x |
| 9 | 5 | | | x | | x | |
| 10a | 6 | | | | x | | |
| 10b | 6 | | | | x | x | |
| 11 | 5 | | | | x | | x |
| 12 | 7 | x | | x | x | | |
| 13 | 1 | x | | x | x | | x |
| 14 | 2 | x | | x | x | x | |
| 15a | 5 | x | | | | | |
| 15b | 5 | x | | | | x | |
| 15c | 5 | x | | | | | x |

Figure 7: Listing of possible merging arrangements and corresponding index of unique merge types for an arbitrary left ($l$) and right ($r$) tree. Note that there are fewer unique types than the total number of possible mergings. The label $l \to v$ ($r \to v$) corresponds to the left (right) tree's variables, and $l \to c$ ($r \to c$) corresponds to the left (right) trees cutpoints. The left and right trees are being merged under the split rule $(v_i, c_i)$. The final two columns indicate whether the left or right tree is actually a terminal leaf node. Note that when a node is a leaf, it cannot have a $v$ or $c$ rule which corresponds to the hashed-out boxes.

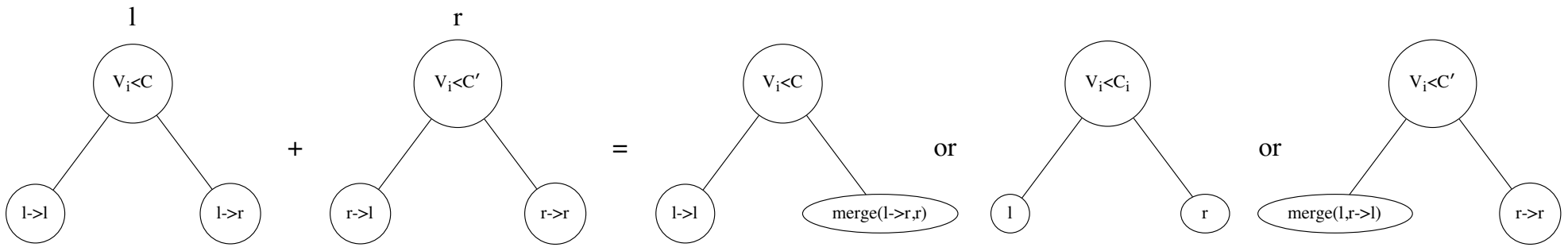

Figure 8: Merging "(+)" possibilities at a given level of the left ($l$) and right ($r$) trees when the current level of the trees both split on the variable $v_i$. This merge type 7 corresponds to arrangement 12 in Figure 7.

where $\pi(T), \pi(T')$ are the prior probabilities of the trees, $L(\cdot)$ represent the integrated likelihoods for the trees, $p_r(T) = \frac{1}{|T|-1}$ is the probability of rotating at a particular internal node of $T$,

$$p_r(T') = \begin{cases} \frac{1}{|T'|-1} & \text{if only 1 way to invert} \\ \frac{2}{|T'|-1} & \text{if two ways to invert,} \end{cases}$$

and $p_m^1$ is calculated as

$$p_m^1 = \begin{cases} \frac{1}{n_m^1 + \mathcal{I}_{\{n_m^1=0\}}+1} & \text{if mergeable} \\ \frac{1}{2} & \text{if leafs} \\ 1 & \text{otherwise,} \end{cases}$$

with similar calculations performed for $p_m^2, p_s^1$ and $p_s^2$. Again, the values of $n_m^1, n_m^2, n_s^1$ and $n_s^2$ are calculated through the recursive application of the merging process, using the merge types outlined in Appendix C and Figure 8. As with other tree proposal mechanisms, the acceptance probability (4) is modified by the constraint requiring all terminal nodes to contain data, leading to automatic rejection if this constraint is not met.

### 4.3 A Simple Rotation Example

A more concrete demonstration of the rotation proposal is shown in Figure 9. Here, the basic structure of the starting tree $T$ is similar to that in Figure 5. Assuming $T_r$ and $T_q$ are not leaf nodes and also do not depend on $X_1$ or $X_2$ (and hence never split or merge), then the behavior of the rotation proposal is simple to follow. In the first rotation (i.e. $T \rightarrow T'$) we will have $p_s^2 = p_m^1 = p_m^2 = 1.0$ and $p_r(T) = \frac{1}{|T_r|+|T_q|+5}$, while $p_s^1 = \frac{1}{3}$ because of the three possible merges of $T_s^L, T_s^R$ in the inverse step. In the second rotation (i.e. $T' \rightarrow T$), we have $p_s^1 = p_s^2 = p_m^2 = 1.0$ and $p_r(T') = \frac{2}{|T_r|+|T_q|+6}$ since the rotation can occur at either node 2 or node 3, and we have $p_m^2 = \frac{1}{3}$ again because of the three possible merges of $T_s^L, T_s^R$.

## 5 Examples

We now return to the two examples introduced in Section 2 to investigate the performance of our proposed sampling strategies. We start with the synthetic example and then explore the Friedman example.

### 5.1 Synthetic Example

In this example, we use the change-of-variable and rotation proposals to allow the sampler to find 3-node and 4-node tree structures that are consistent with the data. Note that in the original posterior found using only birth/death proposals in Section 2.3, only a single 4-node representation was found, simply as a result of the starting random seed (we might have just as easily found a 3-node representation). If we were to augment the birth/death proposal with change, swap or restructure (Wu et al., 2007) proposals, which do not change the tree dimensionality, the diversity of tree structures found would certainly improve, but it is unlikely that the more parsimonious 3-node structure would be sampled if starting from the same random seed. This suggests the rotation proposal greatly increases the diversity of tree structures that can be explored by the MCMC sampler, potentially eliminating the need for random restart or multiple chain approaches to fully explore the posterior.

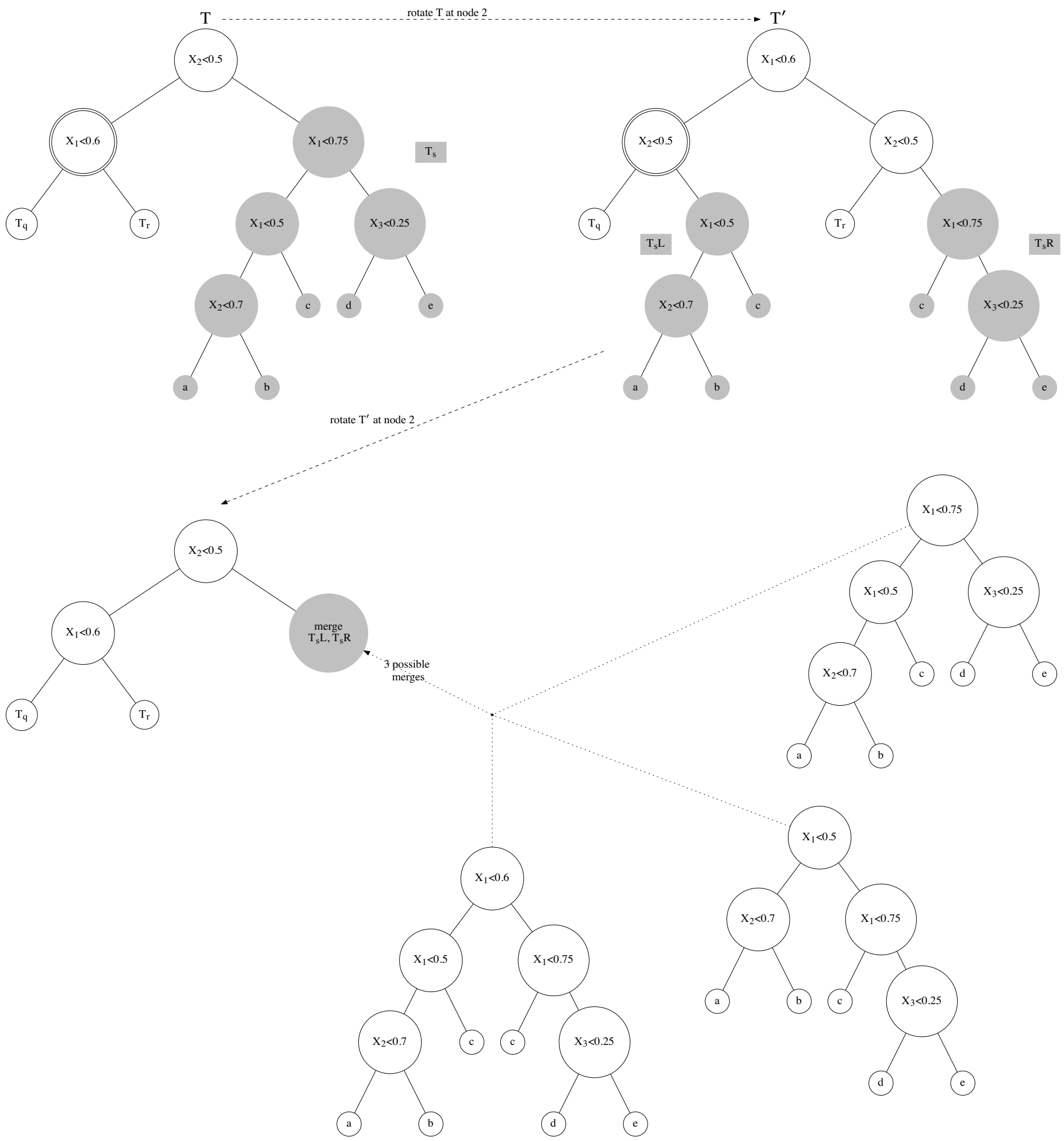

Figure 9: An example of two rotates applied sequentially at the same node (double circled) for a tree similar to the basic structure of Figure 5. Here, the nodes $a, b, c, d$ and $e$ are terminal while $T_q, T_r$ are abitrary trees that do not split on $X_1$ or $X_2$. The tree that will be split and merged, $T_s$, is shown in grey. Note that in the second rotation, the merging type shown in Figure 8 is applied.

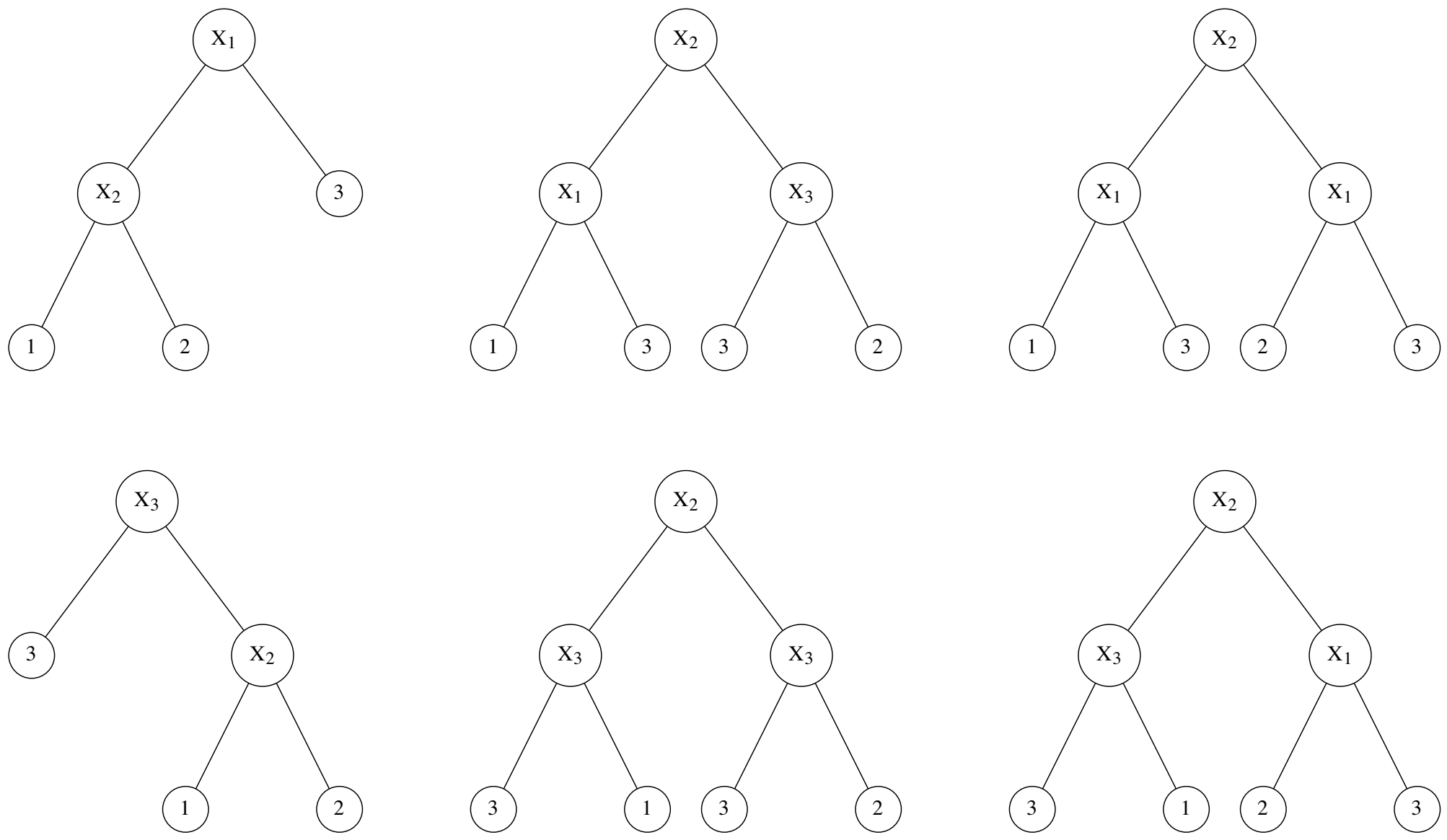

Figure 10: Regression trees found for the synthetic example using the tree rotation and change-of-variable proposal mechanisms using $m = 1$ regression trees. With an additive tree model ($m = 10$), all trees were found using only the tree rotation proposal. The numbers in the leaf nodes indicate the region that maps to that node according to the synthetic function defined in (1). Note that using BART with only birth/death proposals resulted in the sampler converging on the 4-node tree shown in the top-right.

Our preconditioned change-of-variable proposal is also important in this example since $x_1$ and $x_3$ are highly correlated with each other (and independent of $x_2$). Using this proposal mechanisms allows the sampler to propose transitions from $x_1$ to $x_3$ (or vice-versa) about 50% of the time and maintains a high acceptance rate since proposals to transition to $x_2$ are rarely, if ever, made (and would always be rejected if proposed).

The full diversity of trees found in the posterior are shown in Figure 10, which suggests that all possible trees of depth 2 consistent with the data have indeed been sampled when using the proposed change-of-variable proposal and rotation proposal along with the usual birth/death steps with $m = 1$ tree. Forming an additive model with $m = 10$ trees, we were also able to fully explore the possible trees using only the rotation proposal along with birth/death steps.

## 5.2 Friedman Example

In Section 2.2, it was shown that when modeling the Friedman function with BART using only birth/death proposals, the 90% credible interval had a low empirical coverage of around 53% indicating that the posterior was under-representing model uncertainty. It seemed likely that this behavior could be attributed to the poor mixing of the MCMC sampler, which had an acceptance rate for birth/death proposals of only 4%. This low acceptance rate and poor mixing was confirmed by the large negative log integrated likelihood values for these proposals, as shown in Figure 2.

Applying the rotation proposal to the same dataset showed a very noticeable improvement in the behavior of the MCMC sampler, shown in Figure 11. In this case, we selected a rotation step for 20% of the MH proposals, and the usual birth/death step for the remaining 80% of MH proposals. The resulting behavior was a very good acceptance rate of around 25%, and the empirical coverage of the 90% credible interval was around 96%, indicating a conservative coverage with no under-representation of model uncertainty. Looking at the change in log integrated likelihood values for birth/death proposals in the right panel of Figure 11 indicates that the sampler is able to easily explore birth/death proposals that have a good probability of acceptance, even as the rotation step explores different modes of the posterior by altering the internal structure of the regression trees.

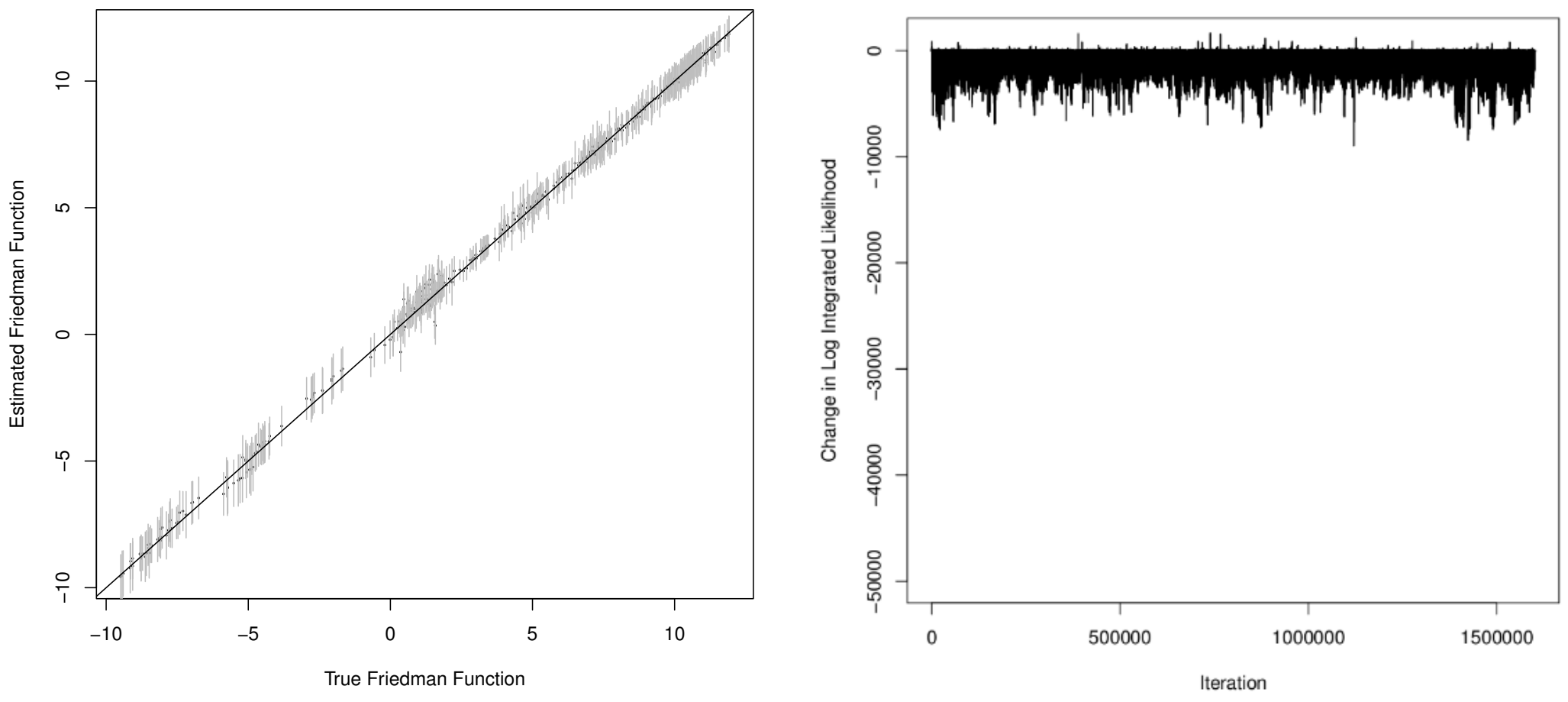

Figure 11: Credible intervals for posterior predictions of the Friedman function with $\sigma^2 = 0.1$ (left) and the change in log integrated likelihood for the birth/death MH proposals (right) when using the rotation proposal. The accuracy of the 90% credible intervals is noticeably improved, having an empirical coverage of 96% and the acceptance rate is 25% indicating good mixing of the MCMC sampler.

In Figure 12, we fit BART to the same data but this time using all MH proposal mechanisms: birth/death proposals, tree rotation proposals and perturb within change-of-variable proposals. In this case, the acceptance rate was 65% and the 90% credible interval had an empirical coverage of 92%. The behavior of the sampler benefits the most from having all three MH steps active, with the proposed changes in log integrated likelihoods for birth/death steps shown in the right panel of Figure 12 confirming the ability of the sampler to easily explore the posterior.

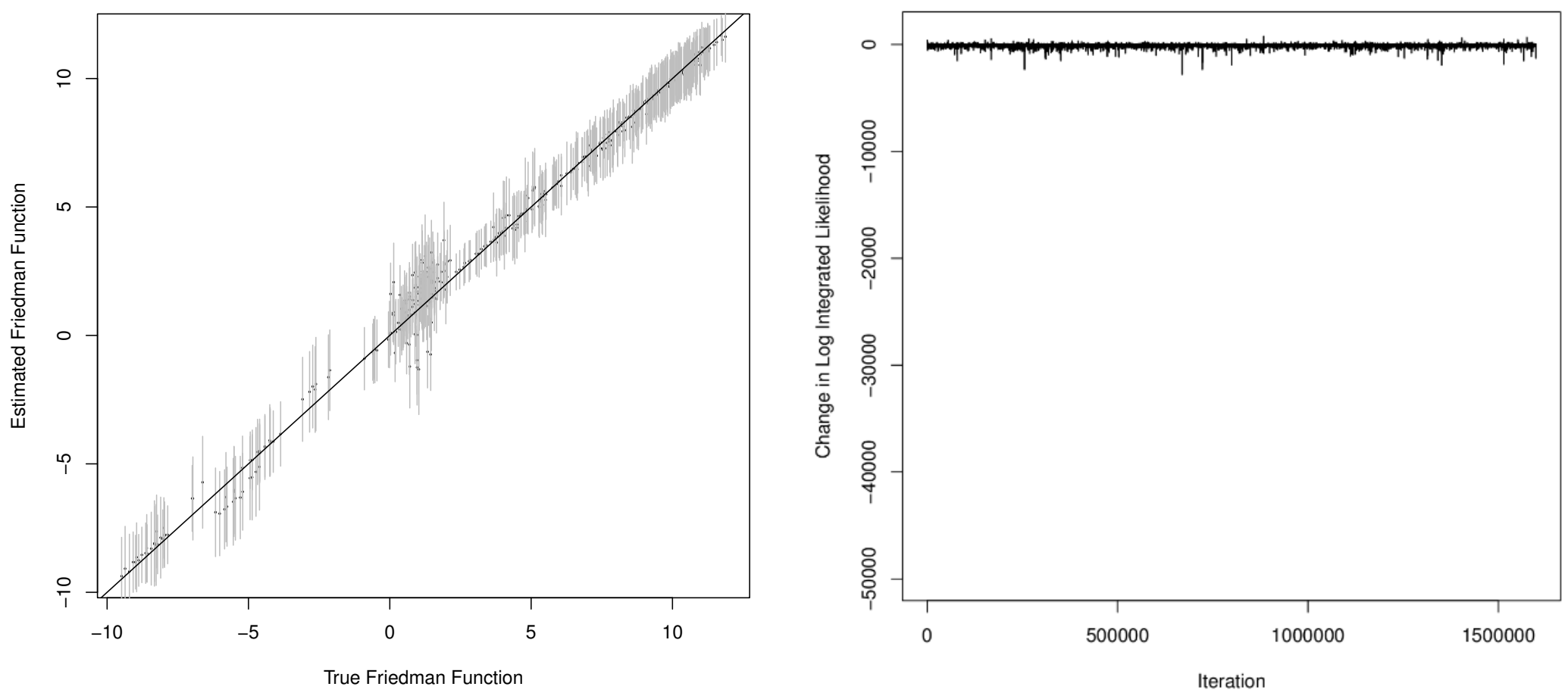

Figure 12: Credible intervals for posterior predictions of the Friedman function with $\sigma^2 = 0.1$ (left) and the change in log integrated likelihood for the birth/death MH proposals (right) when using both rotation and perturb + change-of-variable proposals. The accuracy of the 90% credible intervals is noticeably improved, having an empirical coverage of 92% and the acceptance rate is 65% indicating excellent mixing of the MCMC sampler.

# 6 Discussion

The underlying theme of the MCMC sampling approaches we have discussed is to generate proposals that are consistent with the current state in tree-model space. This avoids needing to use the data to determine which proposals to try while also retaining a high acceptance rate. By consistent, we mean tree arrangements that could only be arrived at by the birth/death process. That is, our proposals could be replaced by a sequence of birth/death moves, but because such moves would have to traverse through a region of low posterior probability to arrive at another mode of high posterior probability, such a sequence does not occur with reasonable probability in practice. This is a key idea underlying all three of the proposed sampling strategies outlined.

The three proposals described try to explore very different sources of information that may account for variability in the posterior distribution. The first is “spatial” variability, explored by our perturbation proposal which proposes new cutpoint values that are consistent with the current tree. The second is the change-of-variable proposal (which necessarily includes a perturbation within it) which explores variability that comes from possible non-independence of covariates. One interpretation of this proposal is of a preconditioned version of the usual change-of-variable proposal, which leads to much higher acceptance rates.

Finally, we explore the structural variability of the tree posterior by using the novel tree rotation proposal to move between tree structures that are nearly equivalent. This method performs local changes to the tree which, over many iterations of the MCMC algorithm, can yield vastly different tree structures in the posterior. It is also not restricted to exploring trees with a fixed number of terminal nodes as the rotation proposal can arrive at trees of larger or smaller dimension.

One interpretation of our rotation proposal is that it is similar to the usual swap proposal except we maintain consistency with the entire tree structure through the described rotation operation. The consistency is enforced by duplicating, cutting and/or merging an appropriate sub-tree during the rotation step.

The approaches described are very different than that of Wu et al. (2007), which essentially tries to explore all three sources of variability in one proposal mechanism. Their approach is very powerful, but for practical purposes is limited by the dimensionality of the covariates.

An advantage of our approach is that the user may decide which proposal mechanisms to use according to their data or questions of interest. For instance, if it is known that the covariates are nearly independent (e.g. by inspection, by performing a Principle Components Analysis (PCA) transformation, or perhaps a variable selection has already taken place), then it is reasonable to dispense with the change-of-variable proposal. If the user is only interested in prediction and does not require the interpretability that comes with fully exploring the tree structure space, then it may be reasonable to dispense with the rotation proposal and rely entirely on the perturbation proposal. Ideally, one might like to use all proposal mechanisms to ensure accurate exploration of the posterior, but in many practical situations a reasonable compromise might be needed.

In the examples we explored in this paper, the proposed methods gave good performance with acceptance rates at least in the low 20% range, and often much higher. For the synthetic example, we used $m = 1$ trees with the change-of-variable and tree rotation proposals which lead to posterior samples that fully explored all possible trees of depth 2. We also noted that with a sum-of-trees representation, the posterior was fully explored using only the tree rotation proposal with $m = 10$ trees.

The Friedman example demonstrated similarly good results. With the tree rotation proposal, the acceptance rate improved from 4% to 25% and the empirical coverage of the 90% credible interval improved from 53% to 96%. Adding in the perturbation proposal saw acceptance rates rise to 65% while the empirical coverage was similarly good at 92%.

In conclusion, the developments in this paper shed new light on ideas for improving the mixing of Bayesian regression tree models in situations where mixing is problematic.

# Appendix A: Generating Right-Rotation Proposal

Listing 1: Pseudocode for generating proposal of a right rotation at node $\eta_i$ splitting on variable $v_i$ at cutpoint $c_i$

```
//keep count of number of tree arrangements and corresponding probabilities
//to be used later in the MH calculation while proposing a rotation
$n_m^1 = 0$, $n_m^2 = 0$, $n_s^1 = 0$, $n_s^2 = 0$
$p_m^1 = 1.0$, $p_m^2 = 1.0$, $p_s^1 = 1.0$, $p_s^2 = 1.0$

//perform rotation at $\eta_i$
rotate_right($\eta_i$)


//cut inadmissible subtrees splitting on $v_i$ after rotation
cut_left(r($\eta_i$),$v_i$,$c_i$)
cut_right(r(p($\eta_i$)),$v_i$,$c_i$)

//calculate nways for splits
if(arenodesleafs(r($\eta_i$), r(r(p($\eta_i$)))) //in this case we could merge the leafs or not
        $p_s^1 = 0.5$
else if(canmerge(r($\eta_i$),r(r(p($\eta_i$))),p($\eta_i$)->v,p($\eta_i$)->c,$n_s^1$)) //calculate $n_s^1$
        $p_s^1$ = 1.0/($n_s^1$+(($n_s^1$==0)?1.0:0.0)+1.0)
else
        $p_s^1 = 1.0$


if(arenodesleafs(l($\eta_i$),l(r(p($\eta_i$)))) //in this case we could merge the leafs or not
        $p_s^2 = 0.5$
else if(canmerge(l($\eta_i$),l(r(p($\eta_i$))),p($\eta_i$)->v,p($\eta_i$)->c,$n_s^2$)) //calculate $n_s^2$
        $p_s^2$ = 1.0/($n_s^2$+(($n_s^2$==0)?1.0:0.0)+1.0)
else
        $p_s^2 = 1.0$


//try merge if not leafs
tmerge = new tree //initialize tmerge as empty tree
p(tmerge)=p($\eta_i$)
if( l(r(p($\eta_i$))) and r(r(p($\eta_i$))) are not leaf nodes ) {
        m = merge( l(r(p($\eta_i$))) , r(r(p($\eta_i$))) , r(p($\eta_i$))->v ,
            r(p($\eta_i$))->c , tmerge, $n_m^1$) //returns a random merged tree in tmerge
        $p_m^1$=1.0/($n_m^1$+(($n_m^1$==0)?1.0:0.0)+1.0)
        double u=gen.uniform()
        if(m and $u > p_m^1$) { //choose tmerge with correct probability
                delete r(p($\eta_i$))
                r(p($\eta_i$)) = tmerge
        }
        else delete tmerged //otherwise we stay with unmerged tree
        if(!m) $p_m^1 = 1.0$ //if no merge possible
}
else {  //both are leafs
        double u=gen.uniform()
        $p_m^1 = 0.5$
        if($u > p_m^1$) { //merge the leafs
                tmerge = a new leaf node
                r(p($\eta_i$)) = tmerge
        }
        else delete tmerge //we don't merge the leafs
}

// Similarly, apply above to $\eta_i$'s left and right subtrees...
tmerge = new tree //initialize tmerge as empty tree
```

```
p(tmerge) = p(η_i)
if( l(η_i) and r(η_i) are not leaf nodes ) {
        m = merge( l(η_i), r(η_i), η_i->v, η_i->c, tmerge, n_m^2)
        u = gen.uniform()
        p_m^2=1.0/(n_m^2+((n_m^2==0)?1.0:0.0)+1.0)
        if(m and u > p_m^2) {
                l(p(η_i)) = tmerge
                delete η_i
        }
        else delete tmerge
        if(!m) p_m^2 = 1.0
}
else {  //both are leafs
        u=gen.uniform()
        p_m^2 = 0.5
        if(u > p_m^2) { //merge the leafs
                tmerge = a new leaf node
                l(p(η_i)) = tmerge
                delete η_i
        }
        else delete tmerge  //we don't merge the leafs
}

//Finally, need a flag to tell us if we can invert this rotation two ways for MH calculation
if(r(p(η_i))->v==η_i->v && r(p(η_i))->c==η_i->c && !isleaf(r(p(η_i))))
        two_ways=true
```

# Appendix B: Cut Algorithm

Listing 2: Pseudocode for left-wise cutting a tree $t$ along variable $v_i$ at cutpoint $c_i$

```
cut_left(tree t, variable v_i, cutpoint c_i)
{
        if(t is not terminal) {
                if(t->v == v_i and t->c >= c_i) {
                        temp = l(t)
                        if(t is left child of p(t) ) {
                                l(p(t)) = temp
                                p(temp) = p(t)
                        }
                        else //t is right child of p(t)
                        {
                                r(p(t)) = temp
                                p(temp) = p(t)
                        }
                        delete t
                        t = temp
                        cut_left(t,v_i,c_i)
                }
                else {
                        cut_left(l(t),v_i,c_i)
                        cut_left(r(t),v_i,c_i)
                }
        }
}
```

# Appendix C: Unique Merge Types

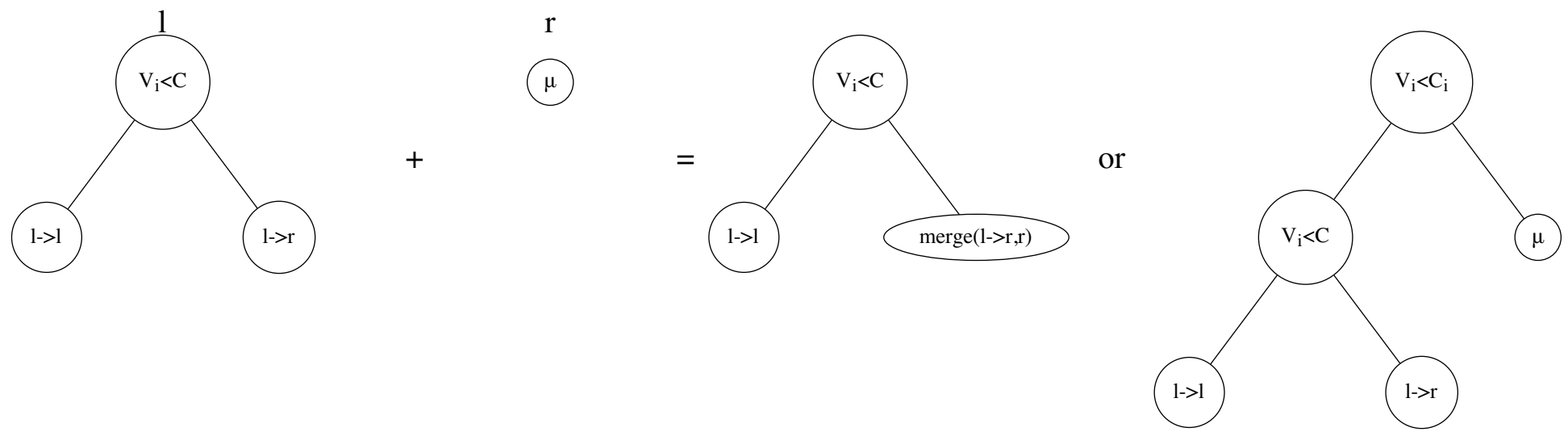

Figure 13: Merging type 1.

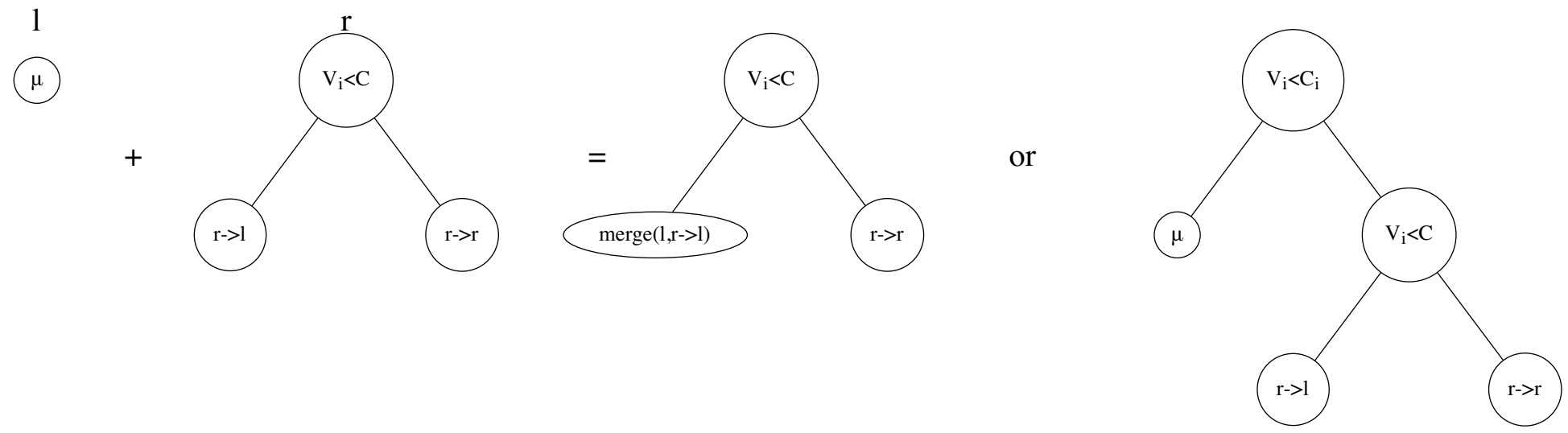

Figure 14: Merging type 2.

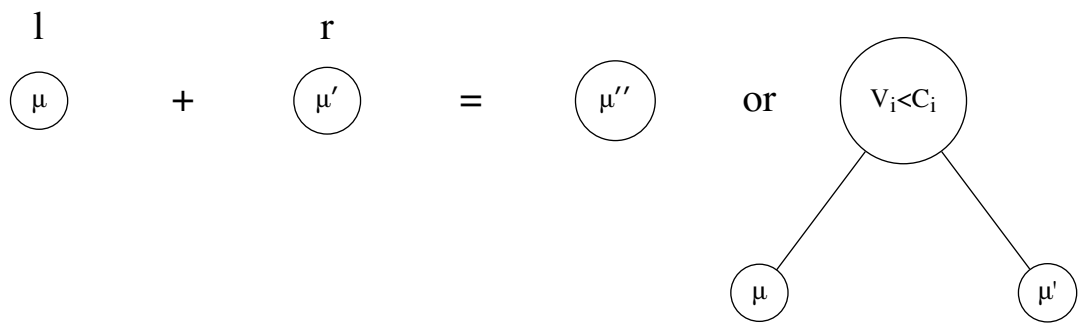

Figure 15: Merging type 3.

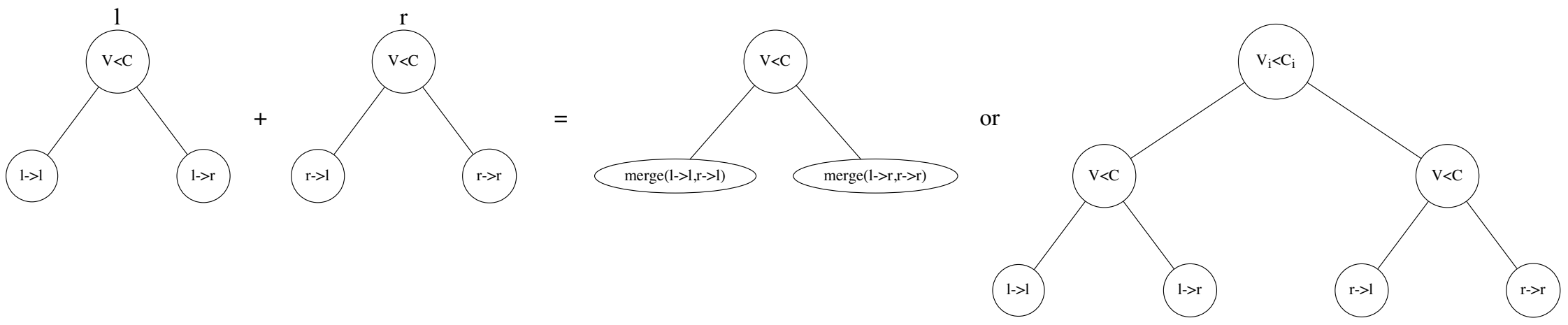

Figure 16: Merging type 4.

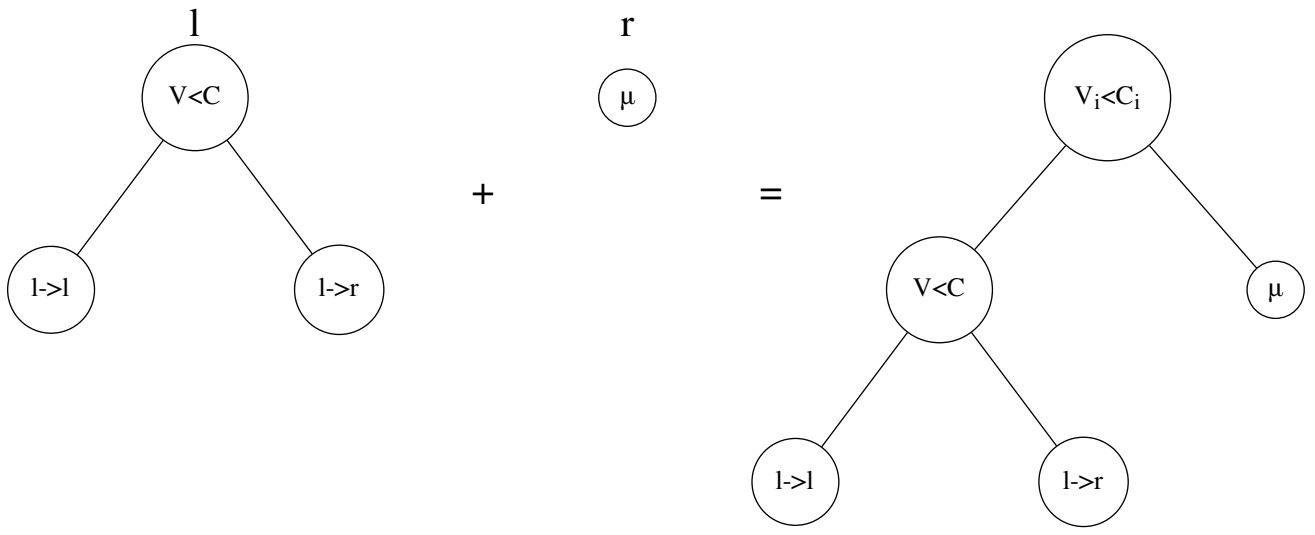

Figure 17: Merging type 5.

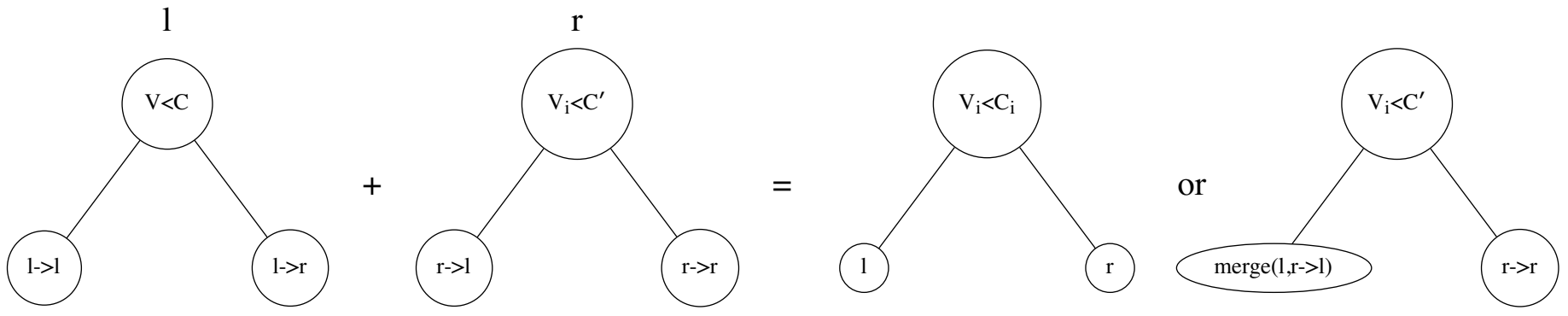

Figure 18: Merging type 6.

# Appendix D: Possible Trajectories Of Right-Rotation

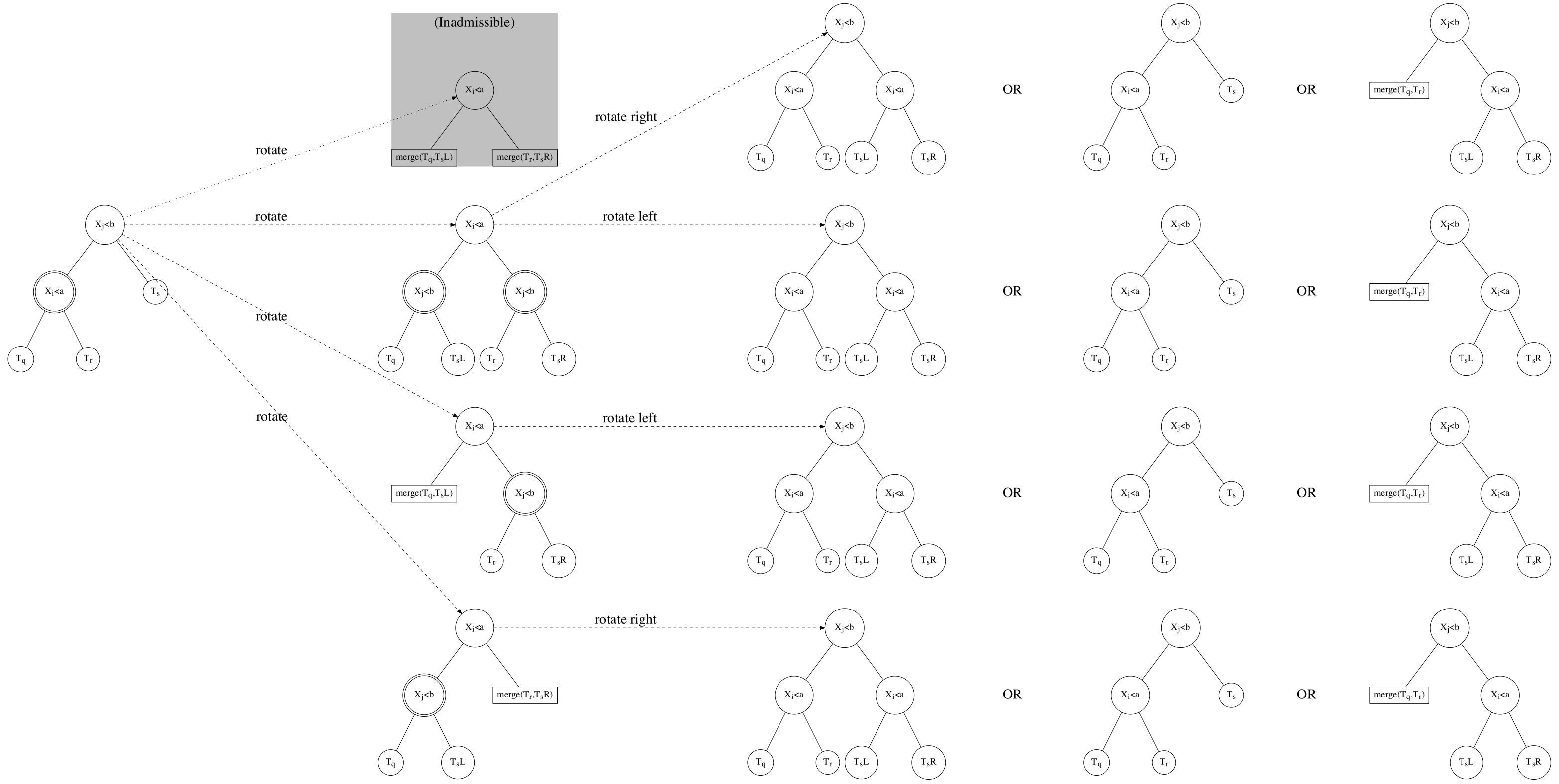

Figure 19: Possible trajectories of an initial right-rotation followed by a subsequent right or left rotation to invert the initial transformation. We denote the non-trivial merges as $merge(\cdot, \cdot)$ which are merges that do not retain the original (variable,cutpoint) being merged on. Note that a non-trivial merge at both the left and right sub-trees results in an inadmissible state since it cannot be inverted to return to the original tree. All other paths show that one can arrive back at the original tree structure, as seen in the fourth column. A similar set of trajectories can be made for an initial left-rotation.